\documentclass[pra,aps,reprint,showpacs,amsmath,amssymb,groupedaddress,longbibliography,nofootinbib]{revtex4-1}
\usepackage{color, graphicx}
\usepackage{hyperref}
\usepackage{cleveref}
\usepackage{amsmath, amssymb, amsfonts}
\usepackage{mathtools}
\usepackage{subfigure}
\usepackage{natbib} 
\newcommand{\cL}{\mathcal{L}}

\newcommand{\cH}{\mathcal{H}}
\newcommand{\cA}{\mathcal{A}}

\newcommand{\bR}{\mathbf{R}}
\newcommand{\br}{\mathbf{r}}
\newcommand{\bx}{\mathbf{x}}

\newcommand{\bv}{\mathbf{v}}

\newcommand{\bk}{\mathbf{k}}

\newcommand{\td}{2\mathrm{D}}
\newcommand{\eff}{\mathrm{eff}}
\newcommand{\intac}{\mathrm{int}}

\newcommand{\ptl}{\partial}

\newcommand{\hphi}{\hat{\phi}}
\newcommand{\hpsi}{\hat{\psi}}
\newcommand{\hvphi}{\hat{\varphi}}

\usepackage{color}

\begin{document}
\baselineskip=0.45 cm

\title{{Roton entanglement in quenched  dipolar Bose-Einstein condensates}}

\author{Zehua Tian}
\affiliation{Seoul National University, Department of Physics and Astronomy, Center for Theoretical Physics, Seoul 08826, Korea}

\author{Seok-Yeong Ch\"{a}}
\affiliation{Seoul National University, Department of Physics and Astronomy, Center for Theoretical Physics, Seoul 08826, Korea}

\author{Uwe R. Fischer}
\affiliation{Seoul National University, Department of Physics and Astronomy, Center for Theoretical Physics, Seoul 08826, Korea}

\begin{abstract}

We study quasi-two-dimensional dipolar Bose-Einstein condensates, in which the Bogoliubov excitation spectrum displays, at sufficiently large gas density, a deep roton minimum 
due to the spatially anisotropic behavior of the dipolar two-body potential.
A rapid quench, performed on the speed of sound of excitations propagating on the 
condensate background, leads to the dynamical Casimir effect, which can be characterized by measuring the density-density correlation function. 
It is shown, for both zero and finite initial temperatures, that the  continuous-variable bipartite quantum state of the created quasiparticle pairs with opposite momenta, resulting from the quench, displays an enhanced potential for the presence of entanglement {(represented by nonseparable and steerable quasiparticle states),}  when compared to a gas with solely repulsive contact interactions. 
{Steerable quasiparticle pairs} contain momenta from close to the roton, and hence {quantum correlations} significantly increase in the presence of a deep roton minimum.  

\end{abstract}

\baselineskip=0.45 cm
\maketitle
\newpage
\section{Introduction}
Quantum field theory predicts that pairs of correlated particles are created from the vacuum 
when the classical background rapidly varies in time \cite{Quantumfield}. 
This process can take place in the expanding (or contracting) 
universe, where it is coined cosmological particle production 
\cite{Schroedinger,PhysRevLett.21.562}, and occurs 
analogously in the dynamical Casimir effect for photons generated 
from the electrodynamical quantum vacuum in a vibrating cavity \cite{DynCasimir}.  
Correlated pairs of particles are also created by the phenomenon of 
Hawking radiation in the presence of an event horizon \cite{Hawking1975,Visser1998}.

{The pairs produced by temporal variations of a homogeneous background consist of quanta with opposite momenta 
and form (continuous-variable) bipartite quantum states.} 
Directly observing pair creation in relativistic quantum field theory is notoriously difficult due to the 
challenging experimental requirements for achieving sizable pair production rates.
To render pair creation, under rather general conditions, accessible to experiment, the idea of quantum simulation 
\cite{QS} was applied to relativistic quantum fields on effective curved spacetime \cite{Unruh,FischerVisser}.
This is frequently classified under the notion of ``analogue gravity," see Ref.~ \cite{BLV} for an extensive review and a comprehensive list of references.  
Several quantum simulation experiments, in which quasiparticles propagate on a rapidly changing background, leading to the dynamical Casimir effect (or analogue cosmological particle production when the induced spacetime metric has a cosmological form), have been proposed, e.g., in 
\cite{ CPP,Schaetz,Carusotto2010,PhysRevD.95.065020,PhysRevLett.103.147003,PhysRevD.95.125003} 
and experiments have been conducted, cf., e.g., \cite{DCE, Lahteenmaki,PhysRevLett.109.220401,Chin}. 
In the same vein, to investigate analogue event and cosmological horizons and the associated effects, several experiments have been proposed \cite{MattCQG,Schuetzhold,Balbinot,PhysRevD.92.024043,PhysRevLett.103.087004, NJP, PhysRevD.96.045012} and some were realized in the lab \cite{PhysRevLett.114.036402,PhysRevLett.105.240401,BHlaser,QCHawking,Gretchen}.

We study in what follows the quantum field theoretical phenomenon of 
pair creation in a 
quasi-two-dimensional (quasi-2D) 
Bose-Einstein condensate (BEC)  with dipolar interactions.
Subjecting the dipolar BEC to rapid temporal changes (quenches) of the condensate background, 
we investigate the production of pairs of quasiparticles and their quantum correlations. 
To assess, then, whether nonseparability and steerability of quasiparticle excitations are present, 
we employ the density-density correlations created by the quench \cite{Finazzi,PhysRevD.96.045012}. 

Magnetic dipole-dipole  interaction (DDI) dominated condensates \cite{Baranov} have been realized with chromium \cite{Chromium}, dysprosium \cite{PhysRevLett.107.190401}, and erbium \cite{PhysRevLett.108.210401} atoms, and the realization of BECs made up of molecules with permanent 
electric dipoles \cite{doi:10.1021/cr300092g} is now at the forefront of ongoing research cf., e.g., \cite{Guo,Rvachov}.
For the creation of quasiparticle pairs in a time-dependent background, we will demonstrate 
that  the existence of a deep {\em roton minimum} in the excitation spectrum \cite{PhysRevLett.90.250403, PhysRevA.73.031602, PhysRevLett.98.030406,PhysRevA.90.043617,PhysRevLett.118.130404} plays a dominant role. 
Various ramifications of the dipolar BEC roton, originally defined for and observed in the strongly interacting 
superfluid helium II \cite{Landau,Mezei}, have been recently experimentally investigated in ultracold dipolar quantum gases \cite{DDIexperiment, PhysRevLett.116.215301,Chomaz,Wenzel}. We will argue below that 
quantum correlations, here represented by the nonseparability and steerability  
present in a bipartite continuous variable system,  
are significantly enhanced in the presence of a deep roton minimum, that is, for sufficiently 
large densities of a DDI dominated gas. 

In quantum simulation--analogue gravity language, 
we study the quasiparticle production due to the dynamical Casimir effect 
for a (in the low-momentum corner) relativistic quantum field (the phonon).  
In the ultracold quantum gas, the shape of the (analogue) {Planckian, Lorentz-invariance-breaking, 
large-momentum sector of the spectrum around  the roton minimum is well controlled. }
We exploit in what follows that the analogue {Planckian} sector can be {\em engineered}, to explore the consequences for the quantum many-body state of the quasiparticles created by a quench. 


\section{Pair creation of quasiparticles in a quasi-2D dipolar gas} \label{section2}
\subsection{Scaling transformation}
We consider an interacting Bose gas comprising atoms or molecules with mass $m$. Its Lagrangian density is given by
($\hbar =1$)
\begin{eqnarray}
\nonumber
\cL&=&\frac{i}{2}(\Psi^\ast\partial_t\Psi-\partial_t\Psi^\ast\Psi)-\frac{1}{2m}|\nabla\Psi|^2-V_\mathrm{ext}|\Psi|^2
\\
&&-\frac{1}{2}|\Psi|^2\int\,d^3\bR^\prime\,V_{\rm int}(\bR-\bR^\prime)|\Psi(\bR^\prime)|^2.
\end{eqnarray}
In the above, $\bR=(\br, z)$ are spatial $3\mathrm{D}$ coordinates. The system is trapped by an external potential of the form $V_\mathrm{ext}(\bR, t)=m\omega^2\br^2/2+m\omega^2_zz^2/2$, where both $\omega$ and $\omega_z$ can in general be time-dependent. We will assume that 
over the whole time evolution, the gas is strongly confined in $z$ direction, with aspect ratio $\kappa=\omega_z/\omega\gg1$. 
We also assume quasi-homogeneity in the plane, i.e.~that the relevant wavelengths of quasiparticle excitations are much shorter than the inhomogeneity scale caused by the in-plane harmonic trapping. 

The two-body interaction is given by
\begin{eqnarray}
V_{\intac}(\bR-\bR^\prime)=g_c\delta^3(\bR-\bR^\prime)+V_\mathrm{dd}(\bR-\bR^\prime),
\end{eqnarray}
where $g_c$ is the contact interaction coupling, and {$V_\mathrm{dd}(\bR-\bR^\prime)=3g_d[\big(1-3(z-z^\prime)^2/|\bR-\bR'|^2)/|\bR - \bR'|^3\big]/4\pi$} describes the dipolar interaction with coupling constant $g_d$. {The dipoles are assumed to be polarized along the $z$-direction by an external field.} In general, $g_c$ and $g_d$ can be time-dependent, depending on the protocol of condensate expansion or contraction which is implemented, see below. We denote by $g_{c,0}$ and $g_{d,0}$ their initial, $t=0$, values. 

To ensure stability in the DDI dominated regime \cite{PhysRevA.73.031602}, we impose that the system remains in the quasi-2D regime during its whole temporal evolution. 
In $z$ direction, we thus assume that the condensate density is a Gaussian, 
$\rho_z(z)=(\pi\,d^2_z)^{-1/2}\exp\big[-z^2/d^2_z\big]$,
where $d_z=b(t)d_{z,0}$ with $d_{z,0}=1/\sqrt{m\omega_{z,0}}$ and $b(t)$ the scale factor \cite{PhysRevLett.118.130404}. {We can then integrate
out the $z$ dependence and obtain the effective quasi-2D interaction, which is given by
$V^{\td}_{\intac}(\br-\br^\prime)=\int\,dzdz^\prime\,V_{\intac}(\bR-\bR^\prime)\rho_z(z)\rho_z(z^\prime)$.} 

Under the usual scaling transformation \cite{PhysRevA.54.R1753, PhysRevLett.77.5315}, laid down in a very general form
in \cite{1367-2630-12-11-113005}, which is applicable to BECs with both time-dependent trapping and coupling constants, one imposes that the {\em scaling variables} $\bx$, $\tau$, and $\psi$ obey  
\begin{eqnarray}\label{ST}
\nonumber
\bx&=&\frac{\br}{b(t)},~~~~~
\tau=\int^t_0\frac{dt^\prime}{b^2(t^\prime)},
\\
\Psi(\br, t)&=&\exp\left[i\frac{mr^2}{2}\frac{\partial_t\,b}{b}\right]\frac{\psi(\bx, \tau)}{b}.
\end{eqnarray} 
{We introduce a factor $f^2=f^2(t)$ in the following,  also see the Heisenberg Eq.~\eqref{Heisenberg} below.  It encapsulates the effects of time-dependent trapping frequency and coupling in the following equation of motion for the scale factor $b$  \cite{1367-2630-12-11-113005,PhysRevLett.118.130404}
\begin{eqnarray}\label{f-square}
\frac{b^3\partial^2_tb+b^4\omega^2(t)}{\omega^2_0}=\frac{g_c(t)}{g_{c, 0}b}=\frac{g_d(t)}{g_{d, 0}b}\eqqcolon f^2(t).
\end{eqnarray}
{Given experimentally prescribed time dependences of trapping and couplings, the above relation determines the scaling expansion.  
On the other hand,} given a desired scaling expansion or contraction $b=b(t)$, to which, e.g., the 
speed of sound $c=c(t)$ time dependence is related by Eq.~\eqref{speed} below via $f(t)$, 
one can determine  
the required trapping frequencies, imposing possibly in addition a temporal dependence of the coupling constants. 
Note that for the scaling approach to accurately yield the expansion or contraction dependence of the field operator in a gas with both contact and
dipolar interactions present (i.e., for the scaling evolution to follow a symmetry), 
the contact ($g_c$) and dipole ($g_d$) couplings are required to either have an identical time dependence, or to both remain constant. 
We remark that when one of the $g_{c,0}$, $g_{d,0}$ equals zero, the
terms ${g_c(t)}/{g_{c, 0}b}$ or ${g_d(t)}/{g_{d, 0}b}$, respectively, 
do not appear as a constraint in 
the equation \eqref{f-square} for 
$f^2$.}

With the above definitions, the Heisenberg equation of motion for the scaling 
field operator $\hpsi(\bx,\tau)$ reads
\begin{eqnarray}
\nonumber
i\ptl_\tau\hpsi &=& \biggl[ - \frac{1}{2m}\nabla_\bx^2 + f^2 \frac{m}{2}\omega^2_0\bx^2 
\\
&&+ f^2 \int d^2\bx' \,V_{\intac,0}^{\td}(\bx - \bx')\hpsi^\dag(\bx')\hpsi(\bx')\biggr]\hpsi .
\label{Heisenberg} 
\end{eqnarray}
with 
$\omega_0 =\omega(t=0)$ the initial trapping frequency.

The quasi-2D dipole-dipole scaling interaction is Fourier transformed according to 
$V_{\intac,0}^{\td}(\bk) =    
\int d^2 \bx e^{-i\bk \cdot \bx} V_{\intac,0}^{\td} (\bx)  $. 
 {We set the (initial) normalization area of the plane to unity in the definition of Fourier transforms and their inverse.  
Also, here and in what follows $\bk$ represents {\em comoving (scaling)} momentum, as we work in the scaling frame of reference.} 
The Fourier transform of the interaction is 
obtained to be 
\cite{PhysRevA.73.031602} 
\begin{equation}
V_{\intac,0}^{\td}(k) = g_0^{\eff}\left( 1 - \frac{3R}{2}\zeta w\left[\frac{\zeta}{\sqrt{2}}\right] \right),
\label{quasi2D_V}
\end{equation} 
where $w[z]=\exp[z^2](1-\mathrm{erf}[z])$ denotes the $w$ function and 
$\zeta = k d_{z,0}$ is a dimensionless wavenumber. Here, we defined an effective contact coupling 
\begin{equation} \label{effective-cc}
g_{0}^{\rm eff}=\frac{1}{\sqrt{2\pi}d_{z,0}}(g_{c,0} + 2g_{d,0})
\end{equation}
and the dimensionless ratio 
\begin{equation}\label{defR}
R= \frac{\sqrt{\pi/2}}{1 + g_{c,0}/2g_{d,0}}.
\end{equation}
The parameter $R$ 
ranges from $R = 0$ if $g_{d,0}/g_{c,0} \to 0$, to $R = \sqrt{\pi/2}$ for $g_{d,0}/g_{c,0} \to \infty$ and expresses the relative strength of contact and dipolar interactions. In the remainder of the paper, 
we put either $R=0$ (contact {dominance}) or $R=\sqrt{\pi/2}$ (DDI {dominance}).

\subsection{Bogoliubov-de Gennes equation}
We decompose the field operator as follows, 
$$\hpsi = \psi_0(1 + \hphi),$$ 
where {$|\psi_0(\bx, \tau)|^2=\rho_0$} represents the condensate density,    and where 
$\hphi$ describes the perturbations (excitations) on top of the condensate. 
The Bogoliubov-de Gennes equation obeyed by the fluctuation field 
$\hphi$ reads  \cite{2001camw.book....1C}
\begin{equation} \label{eqn:20170626_1}
i \ptl_\tau \hphi = \mathcal{H}\hphi + \mathcal{A}(\hphi + \hphi^\dag),
\end{equation} in which we define two operators  
$\mathcal{H}$, $\mathcal{A}$ 
by  
\begin{align} 
\cH & = -\frac{1}{2m}\nabla_\bx^2 - \frac{1}{m \sqrt{\rho_0}} (\nabla_{\bx} \sqrt{\rho_0})\cdot \nabla_{\bx} - i \bv_{\mathrm{com}}\cdot \nabla_{\bx} , \label{eqn:20170626_2}\\
\cA F& = f^2\int d^2\bx'\, V_{\intac,0}^{\td}(\bx - \bx') |\psi_0(\bx')|^2F(\bx'). \notag
\end{align}
where $\cA$ acts by convolution on an arbitrary function $F(\bx)$. 
Here {$\mathbf{v}_{\mathrm{com}}=\frac{1}{m}\nabla_\bx\theta_0$}, {where {$\psi_0 = \sqrt{\rho_0} e^{i\theta_0}$}}, denotes the comoving frame velocity \cite{PhysRevLett.118.130404}.

Assuming vanishingly small comoving velocity, $\mathbf{v}_{\mathrm{com}} = 0$, 
and quasi-homogeneity, $\nabla_\bx \sqrt{\rho_0} \simeq \mathbf{0}$,  {then $\rho_0$ and $\theta_0$ become
independent of $\bx$,}
and we obtain
\begin{eqnarray} \label{eqn:20170614_1}
\nonumber
i \ptl_\tau \hphi &=& -\frac{1}{2m} \nabla_\bx^2\hphi \nonumber\\
& & + f^2 \rho_0 \int d^2 \bx' \: V_{\intac,0}^{\td} (\bx - \bx') (\hphi(\bx') +\hphi^\dag(\bx')). \nonumber\\
\end{eqnarray}
In momentum space, we decompose the fluctuations into their Fourier components,  
$\hphi(\bx) = (1/\sqrt{N}) \sum_{\bk} e^{i\bk \cdot \bx} \hphi_\bk$, $\hphi_\bk = \sqrt{N} 
\int d^2 \bx e^{-i\bk \cdot \bx} \hphi(\bx)$ with $N$ being the total number of atoms in the condensate.

We then have the Fourier space Bogoliubov-de Gennes equation 
\begin{equation} \label{eqn:20170614_3}
i \ptl_\tau \begin{bmatrix} \hphi_\bk \\ \hphi_{-\bk}^\dag \end{bmatrix} = \begin{bmatrix} \cH_\bk + \cA_\bk & \cA_\bk \\ -\cA_\bk & -(\cH_\bk + \cA_\bk) \end{bmatrix} \begin{bmatrix} \hphi_\bk \\ \hphi_{-\bk}^\dag \end{bmatrix}.
\end{equation}
Here, single-particle and interaction energy terms respectively read 
\begin{equation}
\cH_\bk = \frac{k^2}{2m }, \qquad \cA_\bk = f^2 \rho_0 V_{\intac,0}^{\td}(k). \label{Ak_def}
\end{equation} 
{To diagonalize \eqref{eqn:20170614_3}, we thus apply a  Bogoliubov transformation with coefficients $u_{\bk}$ and $v_{\bk}$ as follows:
\begin{equation} \label{eqn:20170614_6}
\begin{bmatrix} \hphi_\bk \\ \hphi_{-\bk}^\dag \end{bmatrix} =
\begin{bmatrix} u_{\bk} & v_{\bk} \\ v_{\bk} & u_{\bk} \end{bmatrix}
\begin{bmatrix} \hvphi_{\bk} \\ \hvphi_{-\bk}^\dag \end{bmatrix},
\end{equation}
where $\hphi_\bk$ and $\hvphi_\bk$ respectively represent the original fluctuation operators and the Bogoliubov quasiparticle operators. Also note that the bosonic algebra imposes 
\begin{eqnarray}\label{NC}
u_\bk^2 - v_\bk^2 = 1.
\end{eqnarray}}
Thereby solving the eigenproblem 
of \eqref{eqn:20170614_3}, we obtain 
\begin{equation}
\begin{array}{l} u_\bk \\ v_\bk \end{array} = \frac{\sqrt{\cH_\bk} \pm \sqrt{\cH_\bk + 2\cA_\bk}}{2(\cH_\bk^2 + 2\cH_\bk \cA_\bk )^{1/4}},
\end{equation}
where where the upper and lower signs refer to $u_\bk$ and $v_\bk$, respectively.
Then, $[u_{\bk} \; v_{\bk}]^T$ is the eigenvector with eigenvalue $\sqrt{\cH_\bk^2 + 2\cH_\bk \cA_\bk}$, and $[v_{\bk} \; u_{\bk}]^T$ is the eigenvector with eigenvalue $-\sqrt{\cH_\bk^2 + 2\cH_\bk \cA_\bk}$. 

{In general, the excitation frequencies are 
scaling time dependent, and Eq.~\eqref{eqn:20170614_3} yields 
\begin{equation} \label{eqnomega}
i \ptl_\tau \begin{bmatrix} 
\hvphi_\bk \\ \hvphi_{-\bk}^\dag 
\end{bmatrix} 
= 
\begin{bmatrix} 
\omega_\bk & {+}  i{\ptl_\tau\omega_\bk}/{2\omega_\bk} \\ 
 {+} i{\ptl_\tau\omega_\bk}/{2\omega_\bk}  & -\omega_\bk 
\end{bmatrix} 
\begin{bmatrix} 
\hvphi_\bk \\ \hvphi_{-\bk}^\dag 
\end{bmatrix},
\end{equation}
where the excitation spectrum is given by 
\begin{equation}\label{omega1}
\omega_\bk(\tau)=\sqrt{\cH_\bk^2 + 2\cH_\bk \cA_\bk(\tau) }.
\end{equation}}
{Here, we introduce the parameter,
\begin{equation}\label{speed}
c(\tau)= f(\tau) \sqrt{g_0^{\rm eff} \rho_0/m}=f(\tau) c_0,
\end{equation}
which is the (scaling time dependent) 
speed of sound. It is the slope of the linear, low-$\bk$ part of the dispersion relation \eqref{omega1}}. We may {also} define, in addition to $R$ in \eqref{defR}, 
another dimensionless parameter 
\begin{equation} \label{A-definition}
A= \frac{mc_0^2}{\omega_{z,0}}=\frac{g_0^{\rm eff} \rho_0}{\omega_{z,0}},
\end{equation} 
representing an effective chemical potential  as measured relative to {the (initial) transverse trapping}, 
linear in both the condensate density and the effective contact coupling defined in \eqref{effective-cc}.  

\begin{figure}[t]
\centering
\includegraphics[width=0.4\textwidth]{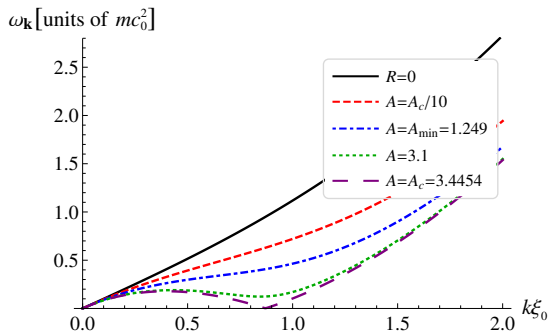}
\caption{(Color online) {\em Stationary state excitation spectrum.}
Bogoliubov excitation energy in units of $mc^2_0$, for DDI {dominance}, $R=\sqrt{\pi/2}$. For $A>A_\mathrm{min}=1.249$, the spectrum develops a roton minimum and becomes unstable for  $A>A_c=3.4454$. $R=0$ denotes the contact interaction case where  the Bogoliubov excitation energy, 
when normalized to $mc_0^2$, as here, is independent of $A$; $\xi_0$ is the healing length defined in\eqref{def_xi0}.}
\label{Omegadd1}
\end{figure}

For a stationary state $f=1$ $[c(\tau)=c_0]$, the {healing length} is given by
\begin{equation}
\xi_0=\frac1{mc_0}. 
\end{equation}\label{def_xi0}
{The inverse of $\xi_0$, $k_{\rm Pl}\coloneqq 1/\xi_0$, is an analogue ``Planck scale."  
Close to the roton minimum at $k\xi_0\approx 0.9$, then, Lorentz invariance is {strongly} broken and a particular variant of 
{\em Planckian} ($k\sim k_{\rm Pl}$) physics can be simulated \cite{PhysRevLett.118.130404}.}
In Fig.\,\ref{Omegadd1}, we plot the corresponding stationary state 
Bogoliubov excitation energy, from which we see 
that the spectrum in a strongly dipolar BEC develops a roton minimum for sufficiently large $A$.
The system becomes unstable past the critical value $A=A_c=3.4454$ 
(when $R=\sqrt{\pi/2}$ \cite{PhysRevA.73.031602}).  
In the low-momentum corner, the spectrum is generally linear in momentum,  
\begin{equation}\label{low-frequency}
\omega_\bk = c_0 k \qquad (k\xi_0\ll 1), 
\end{equation}
implying the (pseudo-)Lorentz invariance of the system from which the effective
metric concept for the propagating quantum field of phonons emerges \cite{BLV}.


For a stationary system, we find that the operators $\hvphi_\bk$ and 
$\hvphi^\dag_{-\bk}$ decouple, and oscillate at constant frequencies $\pm\omega_\bk$, where $\tau=t$ for the stationary case with $f=b=1$ in \eqref{ST}, 
\begin{eqnarray}\label{d-phi}
\hvphi_\bk(\tau)=\hat{b}_\bk\,e^{-i\omega_\bk t},~~~~~~\hvphi^\dag_{-\bk}(\tau)=\hat{b}^\dag_{-\bk}\,e^{i\omega_\bk t}.
\end{eqnarray}
Here $\hat{b}_\bk$ and $\hat{b}^\dag_\bk$ are, respectively, annihilation and creation operators of collective excitations with momentum $\bk$ above the stationary condensate. 

\subsection{Mode mixing}
As a result of a rapid temporal change of $c^2$, as defined in \eqref{speed}, and 
which is encoded in the scale factor $f$ defined in \eqref{f-square}, Eq.~\eqref{eqnomega} engenders mode mixing between the quasiparticle modes of momenta $\bk$ and 
$-\bk$, which entails the amplification of quantum and thermal fluctuations.
It is convenient to characterize the mode mixing by introducing the coefficients $\alpha_\bk(\tau)$ and $\beta_\bk(\tau)$ \cite{PhysRevA.89.063606}:
\begin{eqnarray}\label{ab}
\nonumber
\hvphi_\bk(\tau) 
&=&\left[\alpha_\bk(\tau)\hat{b}^\mathrm{in}_\bk+\beta^\ast_\bk(\tau)\hat{b}^{\mathrm{in}\dagger}_{-\bk}\right]
\exp\left[\textstyle -i\int^\tau\omega_\bk(\tau^\prime)d\tau^\prime\right], \nonumber
\\  
\hvphi^\dagger_{-\bk}(\tau)&=&\left[\alpha^\ast_\bk(\tau)\hat{b}^{\mathrm{in}\dagger}_{-\bk}+\beta_\bk(\tau)\hat{b}^{\mathrm{in}}_\bk\right]\exp\left[\textstyle i\int^\tau\omega_\bk(\tau^\prime)d\tau^\prime\right].\nonumber\\
\end{eqnarray}
In the limit $\tau\rightarrow-\infty$, $\hat{b}^\mathrm{in}_\bk$ and $\hat{b}^{\mathrm{in}\dagger}_{-\bk}$ are defined such that $\hvphi_\bk(\tau)\rightarrow \hat{b}^\mathrm{in}_\bk e^{-i\omega_\bk \tau}$, $\hvphi^\dagger_{-\bk}(\tau)\rightarrow\hat{b}^{\mathrm{in}\dagger}_{-\bk}e^{i\omega_\bk \tau}$, or equivalently, $\alpha_\bk\rightarrow1$ and $\beta_\bk\rightarrow0$ as $\tau\rightarrow-\infty$. That is to say, the operators $\hat{b}^\mathrm{in}_\bk$ and $\hat{b}^{\mathrm{in}\dagger}_{\bk}$ are, respectively, the annihilation and creation operators of collective excitations with momentum $\bk$ in the initial stationary state.  
From Eqs.~\eqref{eqnomega} and \eqref{ab}, we find that the evolution of the operators $\hvphi_\bk(\tau)$ and $\hvphi^\dagger_{-\bk}(\tau)$ is completely determined by $\alpha_\bk(\tau)$ and $\beta_\bk(\tau)$, and the corresponding evolution equations of $\alpha_\bk(\tau)$ and $\beta_\bk(\tau)$ are:
\begin{eqnarray}
\nonumber
\ptl_\tau\alpha_\bk&=& {+}\frac{\ptl_\tau\omega_\bk}{2\omega_\bk}\exp\bigg(2i\int^\tau\omega_\bk(\tau^\prime)d\tau^\prime\bigg)\beta_\bk,
\\
\ptl_\tau\beta_\bk&=& {+}\frac{\ptl_\tau\omega_\bk}{2\omega_\bk}\exp\bigg(-2i\int^\tau\omega_\bk(\tau^\prime)d\tau^\prime\bigg)\alpha_\bk.\label{eqab}
\end{eqnarray}
Given the temporal change $c^2=c^2(\tau)$, the above equations can be solved to obtain $\alpha_\bk(\tau)$ and $\beta_\bk(\tau)$, and hence $\hvphi_\bk(\tau)$ and $\hvphi^\dagger_{-\bk}(\tau)$.

The phase factors of $\alpha_\bk$ and $\beta_\bk$ in \eqref{eqab} determine the phase of the oscillations of the density-density correlation function (i.e., Eq.~\eqref{cf2}) around its mean value.
We note in this regard that a typo has occurred in Eqs.\,(21) and (26) of Ref. \cite{PhysRevD.95.065020}, where the sign ``$-$" should be a ``$+$". {However, this sign has no effect on 
the minimal values of the density-density correlation amplitude, which is characterizing the degree of entanglement present in the quasiparticle state after the quench. We shall now turn  
to the corresponding discussion.}


\section{Characterizing nonseparability and steerability}\label{section3}
{After a series of identical experiments to measure the density distribution of the gas, 
we can extract its mean, as well as  the fluctuations around this mean.} 
The corresponding density-density correlations \cite{Hung} are related to the quasiparticle quantum state \cite{Finazzi}. We will proceed to demonstrate how to use these correlations to measure nonseparability and {steerability} between the created quasiparticles with opposite {momenta} $\bk$ and $-\bk$, which are due to temporal variations of the condensate background. {Below, we closely follow the 
density-density correlation-function based discussion of the criteria for nonseparability and steerability 
previously laid down in Refs.\,\cite{PhysRevD.95.065020, PhysRevD.96.045012}.}

\subsection{The density-density correlation function}

The total atom number density in the condensate is given to 
linear order in the fluctuations by 
\begin{eqnarray}
\nonumber
\hat{\rho}(\tau, \bx)&=& \hpsi^\dagger(\tau, \bx)\hpsi(\tau, \bx)
\simeq\rho_0(1+\hphi^\dagger(\tau, \bx)+\hphi(\tau, \bx)).\\
\end{eqnarray}
In a homogeneous system, the background density $\rho_0$ is constant, and 
the relative density fluctuation is 
\begin{eqnarray}\label{RDF}
\frac{\delta\hat{\rho}(\tau, \bx)}{\rho_0}=\frac{\hat{\rho}(\tau, \bx)-\rho_0}{\rho_0}
=\hphi^\dagger(\tau, \bx)+\hphi(\tau, \bx).
\end{eqnarray}
We consider \emph{in situ} measurements of $\delta\hat{\rho}(\tau, \bx)$ performed at some (scaling) measurement time $\tau=\tau_{\rm m}$.
From the equal-time commutators $[\hpsi(\tau, \bx), \hpsi(\tau, \bx^\prime)]=0$ and 
$[\hpsi(\tau, \bx), \hpsi^\dagger (\tau, \bx^\prime)]=\delta(\bx-\bx^\prime)$,
one can easily verify that $\delta\hat{\rho}(\tau, \bx)$ and $\delta\hat{\rho}(\tau, \bx^\prime)$ commute with each other. 

In momentum space,  
we express  the relative density fluctuation \eqref{RDF} in terms of quasiparticle operators, 
\begin{eqnarray}\label{MRDF}
\frac{\delta\hat{\rho}_\bk(\tau)}{\rho_0}&=&\hphi_\bk(\tau)+\hphi^\dagger_{-\bk}(\tau)\nonumber\\
&=&(u_{\bk}+v_{\bk})(\hvphi_\bk(\tau)+\hvphi^\dagger_{-\bk}(\tau)).
\end{eqnarray}
Note that taking the Hermitian conjugate of the operator \eqref{MRDF} is equivalent to changing the sign of $\bk$, {as a consequence of the fact that the relative density fluctuation operator in \eqref{RDF} is itself a Hermitian operator and thus is an observable (the results of the corresponding measurement are real quantities).} It is straightforward to show that this operator commutes with its Hermitian conjugate, and thus the following correlation function is well defined:
\begin{multline}\label{cf1}
G_{2, \bk}(\tau)=\frac{\langle|\delta\hat{\rho}_\bk(\tau)|^2\rangle}{\rho_0^2} \\
=(u_{\bk}+v_{\bk})^2(2n_\bk+1+2\Re[c_\bk e^{-2i\omega_\bk \tau}]),
\end{multline}
where $n_\bk=\langle \hat{b}^\dagger_\bk\hat{b}_\bk\rangle$ is mean occupation number,
and  $c_\bk=\langle \hat{b}_\bk\hat{b}_{-\bk}\rangle$ is pair amplitude.
To obtain the above relation, {the assumption} 
$n_{\bk}=n_{-\bk}$ has been made. 
 The second line holds when the background has reached a stationary state, so 
 that the frequencies $\omega_\bk$  become time-independent. 

 
 The mean occupation number $n_\bk$ determines the time-averaged mean of $G_{2, \bk}(\tau)$, while 
the magnitude and phase of the correlation $c_\bk$ respectively determine the magnitude and phase of the oscillations of $G_{2, \bk}(\tau)$ around its mean value.
{For the zero temperature (quasiparticle vacuum) case, i.e., $n_\bk=0$ (and hence $c_\bk=0$), in the correlation function there is just one constant term (the ``$+1$") left}, which is measurable as well and encodes the vacuum fluctuations of the quasiparticle field. This will become of importance later on.

For a thermal initial state with the equilibrium distribution $2n_\bk+1=\coth\big({\omega_\bk}/{2T}\big)$, the term containing $c_\bk$ vanishes and the correlation function in Eq.~\eqref{cf1}  reads 
\begin{widetext}
\begin{equation}
\label{corr-f1}
G_{2, \bk}=\frac{k \xi_0/2}{\sqrt{\big(\frac{k \xi_0}{2}\big)^2+1-\frac{3R}{2}k\xi_0\sqrt{A} w\left[\frac{k\xi_0}{\sqrt{2}}\sqrt{A}\right]}}
\coth\left[\frac{k\xi_0\sqrt{\big(\frac{k \xi_0}{2}\big)^2+1-\frac{3R}{2}k\xi_0\sqrt{A} w\left[\frac{k\xi_0}{\sqrt{2}}\sqrt{A}\right]}}{2T/mc^2_0}\right].
\end{equation} 
\end{widetext}
\begin{figure}[hbt]
\centering
\subfigure[]{\includegraphics[width=0.3\textwidth]{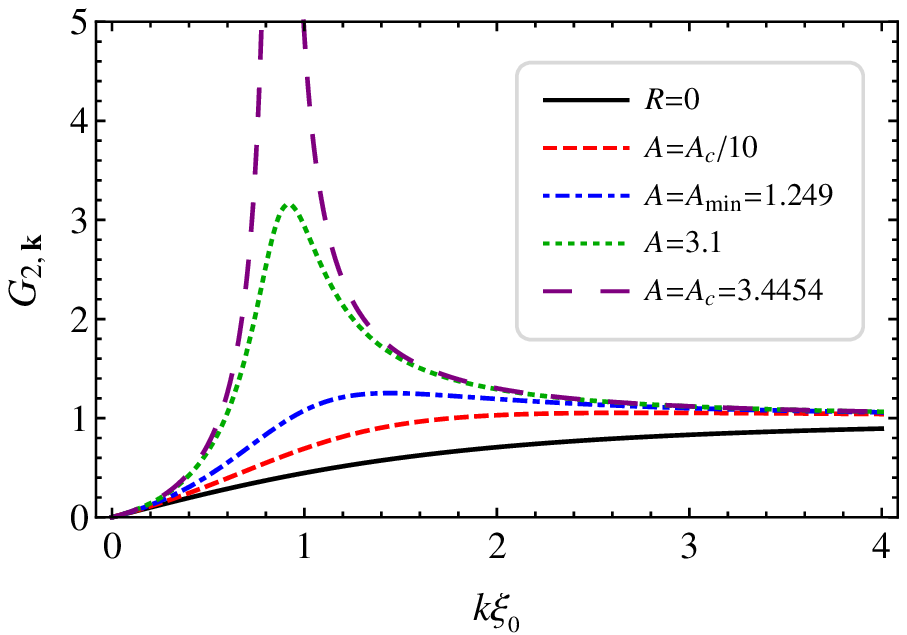}}
\subfigure[]{\includegraphics[width=0.3\textwidth]{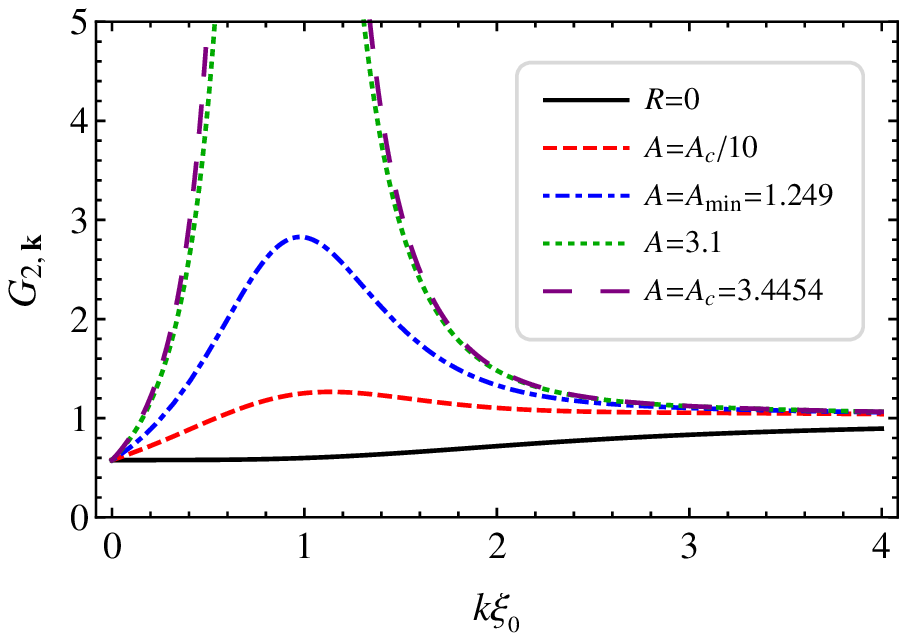}}
\subfigure[]{\includegraphics[width=0.3\textwidth]{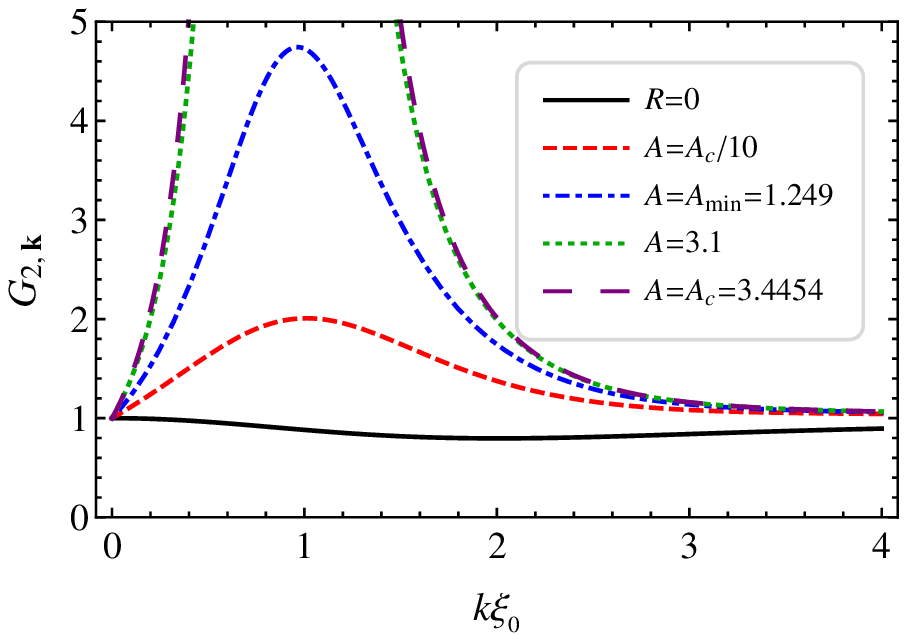}}
\caption{(Color online) {\em Stationary state density-density correlations for increasing 
temperature (from left to right).} 
The density-density correlation function $G_{2, \bk}$ 
in thermal and quasiparticle ground states.
The initial temperatures are (a) $T/mc^2_0=0$, (b) $T/mc^2_0=1/\sqrt{3}$, and (c) $T/mc^2_0=1$.
The black solid line corresponds to contact interaction,
$R=0$ ($A=A_c/10$). DDI dominated cases ($R=\sqrt{\pi/2}$) are shown 
by the remaining curves with $A$ specified in the insets. 
}\label{Gdd}
\end{figure}

In Fig.\,\ref{Gdd}, we plot the thermal density-density correlation function \eqref{corr-f1} 
of a dipolar BEC at various initial temperatures (in units of $mc^2_0$), and as a function of the nondimensionalized momentum $k\xi_0$, with fixed $A$ and $R$. We see that the density-density correlation function is strongly modified near  the roton minimum of the spectrum. In particular, when
the roton minimum approaches zero (near criticality), the modification of the density-density correlation function relative to the pure contact case diverges.

We now discuss the high- and low-temperature limits of \eqref{corr-f1}
separately. When $\omega_{\mathbf k}/T\ll 1$, we have 
$2n_\bk+1=\coth\big({\omega_\bk}/{2T}\big)\simeq2T/\omega_\bk+\omega_\bk/6T$, so that Eq.~\eqref{corr-f1} becomes in the low-momentum (phonon) limit, expanding to quadratic order in $k\xi_0$, 
\begin{eqnarray}
G_{2, \bk}&=&\frac{T}{mc^2_0}\left[1+\frac{3}{2}\sqrt{A}Rk\xi_0\right] \nonumber\\
& & -\left[\frac{T}{mc^2_0}\left(\frac{1}{4}+\frac{3AR}{\sqrt{2\pi}}+\frac{9AR^2}{4}\right)-\frac{mc^2_0}{12T}\right](k\xi_0)^2  \nonumber\\  & & +{\mathcal O} ((k\xi_0)^3).
\end{eqnarray}
For the contact interaction, i.e., $R=0$ case, we reproduce the result of Ref.~\cite{PhysRevD.95.065020}. 
On the other hand, we see that for finite relative strength $R$ and density of dipoles encapsulated in $A$, 
both $R$ and $A$ enter the correlation function.
For $k\rightarrow0$, $G_{2, \bk}$ simply approaches the dimensionless temperature $T/mc^2_0$;  
one can thus determine the temperature of the gas by examining the low-momentum density fluctuations.

When $\omega_\bk/T\gg1$, we have $\coth\big({\omega_\bk}/{2T}\big)\simeq1$, so that Eq.~\eqref{corr-f1} becomes 
\begin{eqnarray}
G_{2, \bk}\simeq \frac{k \xi_0/2}{\sqrt{\big(\frac{k \xi_0}{2}\big)^2+1-\frac{3R}{2}k\xi_0\sqrt{A} w\left[\frac{k\xi_0}{\sqrt{2}}\sqrt{A}\right]}}.
\end{eqnarray}
Again, the difference to the contact case $R=0$ is manifest, because the relative
strength and density of dipoles are explicitly involved via $R$ and $A$, respectively.
In the high-momentum limit of free particles $k\xi_0\gg 1$, $G_{2, \bk}$ 
approaches unity, regardless of temperature and interactions \cite{NoteAsymptote}.

\subsection{Criteria for nonseparability and steerability of quasiparticle pairs}
\label{steersection}

Pair production in a time-dependent background can be caused 
by quasiparticles already present, e.g. in a thermal state, 
or emerge from quasiparticle quantum vacuum fluctuations.
The created pairs possess opposite momenta $\bk$ and $-\bk$ 
and are correlated. To study their correlations, we restrict ourselves to the consideration of bipartite quantum states.

{We focus on quantum correlations represented by entanglement, the notion coined by Schr\"odinger
in 1935 \cite{Schrödinger1935}, which will here be represented by nonseparable and 
steerable quasiparticle states. {\em Steering,} as introduced by Schr\"odinger in
the same year \cite{schršoedinger_1935}, refers to the correlations that can be observed between the outcomes of measurements applied on half of an entangled state (Alice) 
and the resulting post-measurement states that are then left with the other party (Bob). 
A criterion testing quantum steering can be seen as an entanglement test where one of the parties (Alice) performs {\em uncharacterized} 
measurements, i.e. with a procedure not accessible (hidden in a black box) to the other party (Bob) \cite{Cavalcanti}. 
Steerable entangled quantum states are a strict subset of nonseparable states, and a strict superset 
of states exhibiting Bell nonlocality \cite{WisemanPRL,WisemanPRA,Reid}. }

{Criteria to assess the degrees of correlation between the created quasiparticles using density-density correlations have previously been analyzed in detail in Refs. \cite{PhysRevD.95.065020, PhysRevD.96.045012}. 
Nonseparability and steerability of quasiparticle pairs are achieved when}
\begin{eqnarray} \label{ie2}
G_{2, \bk}(\tau)<G_{2, \bk}^\mathrm{vac}=(u_{\bk}+v_{\bk})^2\quad \mbox{[Nonseparable]},
\end{eqnarray}
and 
\begin{eqnarray} \label{QS3}
G_{2, \bk}(\tau)<\frac12 G^\mathrm{vac}_{2, \bk} \quad \mbox{[Steerable]}.
\end{eqnarray}
Here, $G_{2, \bk}^\mathrm{vac}$ is the correlation 
due to quasiparticle vacuum fluctuations.  
Whenever $G_{2, \bk}(\tau)$ dips below its vacuum value for some times, the state is {necessarily nonseparable}.
Compared with the nonseparability condition in \eqref{ie2}, the criterion for steerability shown in \eqref{QS3} is obviously more stringent (as it should be), due to the factor of 1/2 on the right hand side, again reflecting the fact that quasiparticle states exhibiting steering form a subset of nonseparable states.
{We also note here that a concrete experimental protocol to assess quasiparticle entanglement by the covariance matrix of the quasiparticle quadratures was proposed in Ref.~\cite{Finazzi}.}

\section{Dynamical Casimir effect} \label{DCE}
\subsection{Rapid changes of the sound speed}
\begin{figure}[t]
\centering
\includegraphics[width=0.33\textwidth]{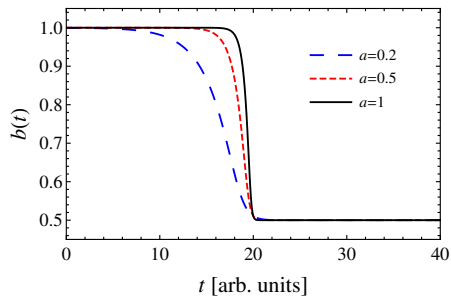}
\caption{(Color online) 
{The scale factor $b(t)$ in \eqref{scale-factor}, as a function of the lab time $t$, 
showing the compression of the condensate for various speed of sound quench rates $a$ (in arbitrary units of inverse time). Here, we take $c^2_i/c^2_f=1/2$.}}
\label{figb}
\end{figure}
\begin{figure*}[t]
\centering
\subfigure[]{\includegraphics[width=0.34\textwidth]{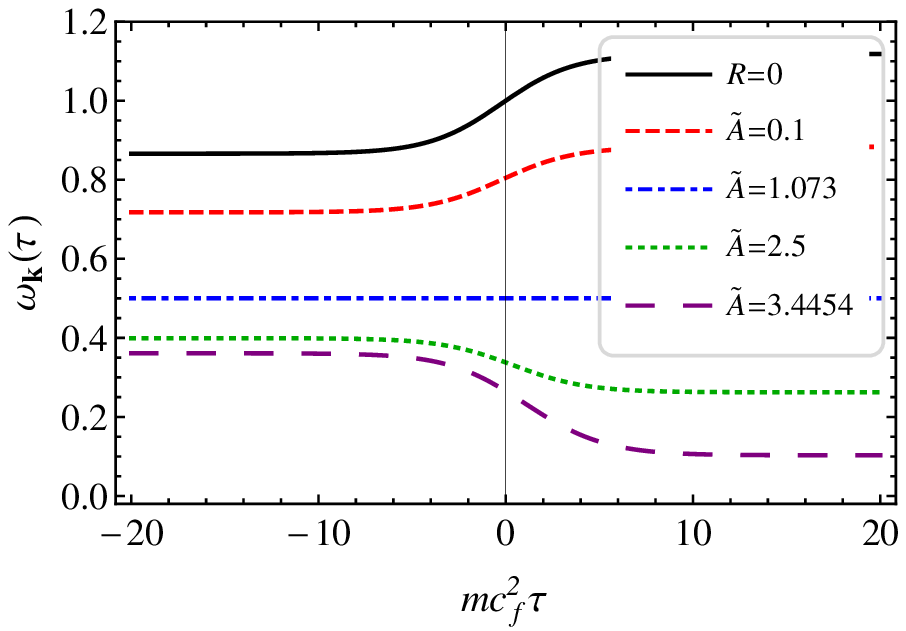}}\hspace*{2em}
\subfigure[]{\includegraphics[width=0.33\textwidth]{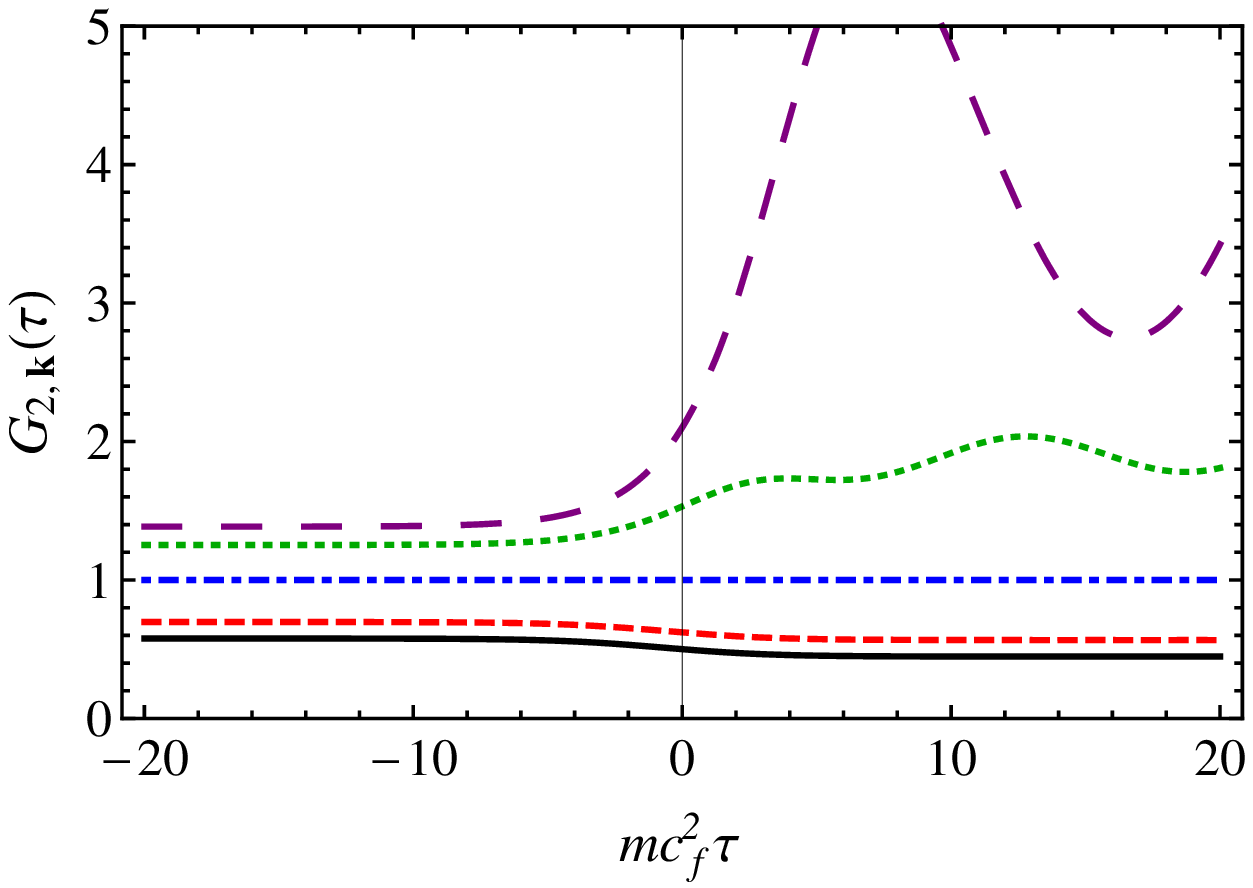}}
\subfigure[]{\includegraphics[width=0.34\textwidth]{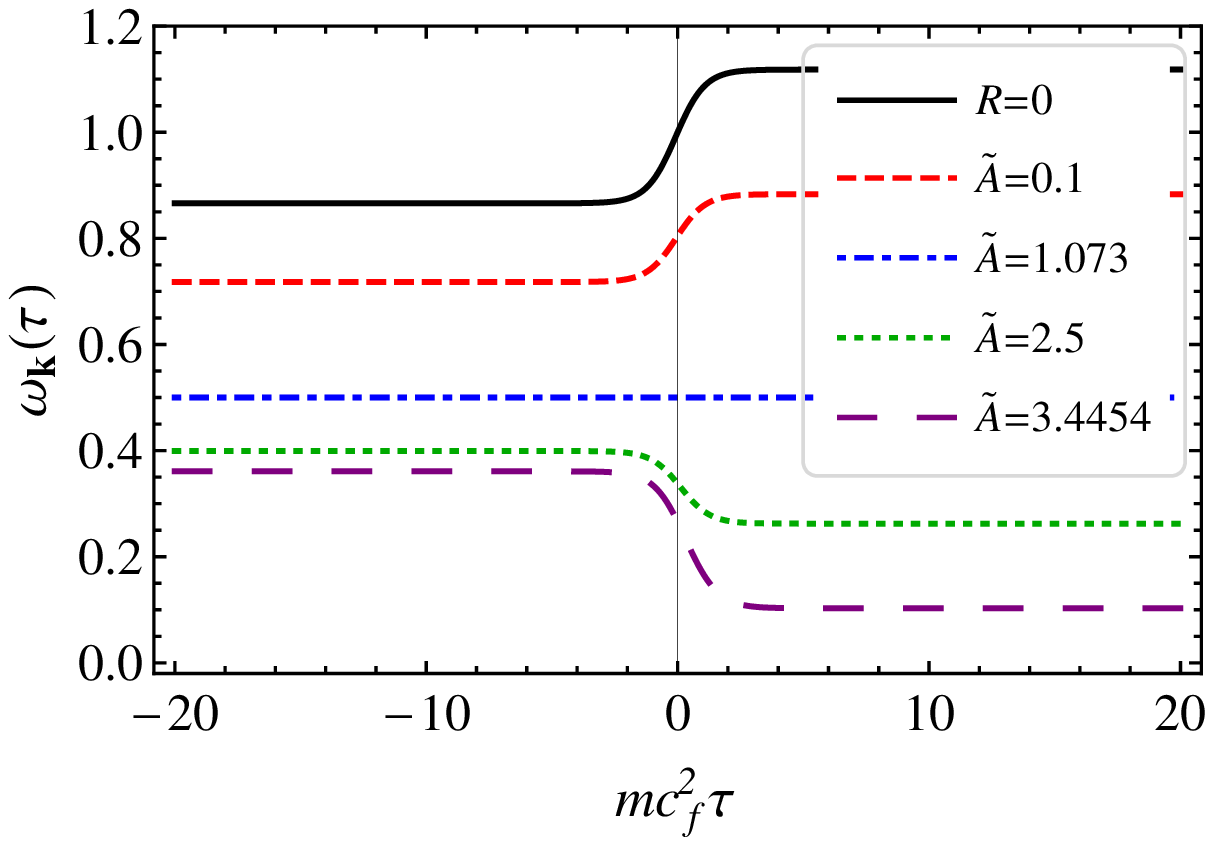}}\hspace*{2em}
\subfigure[]{\includegraphics[width=0.33\textwidth]{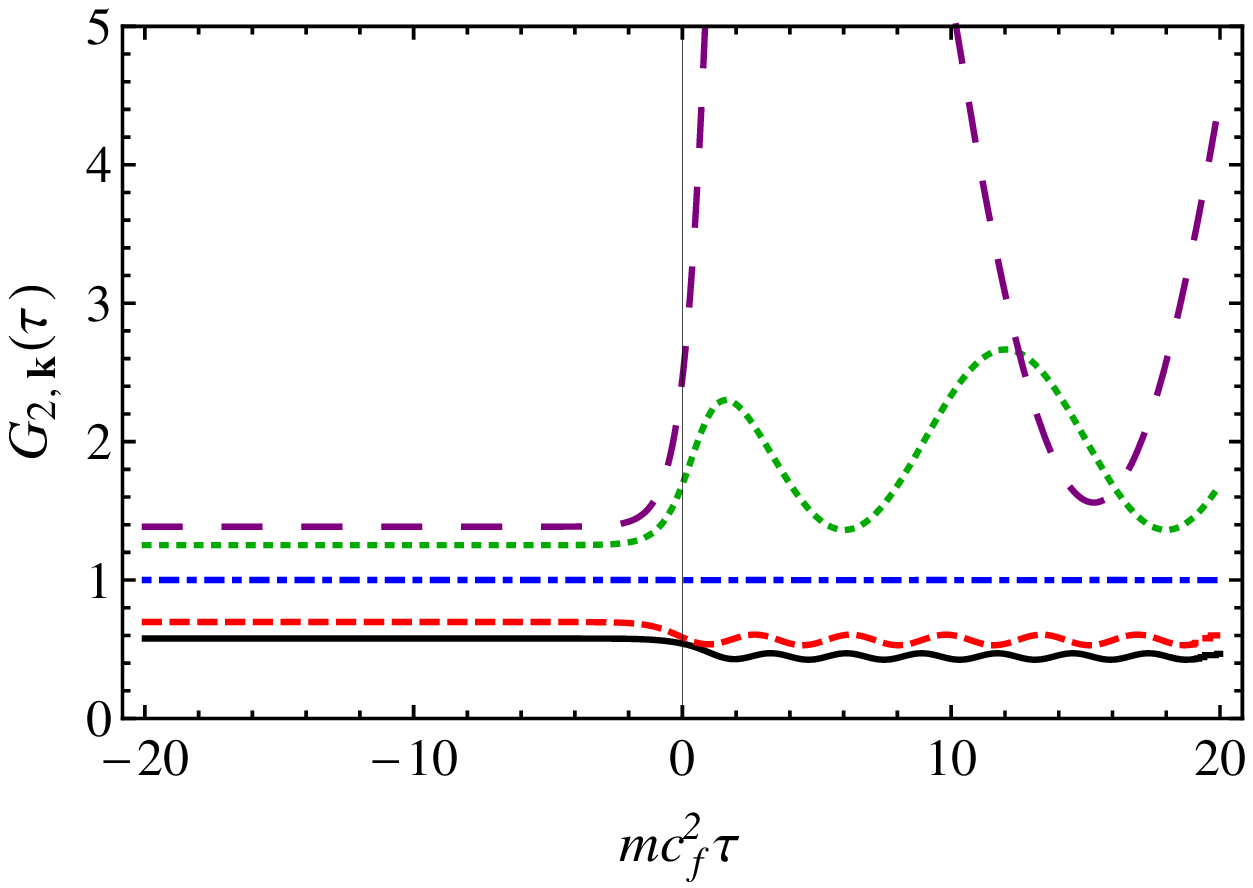}}
\caption{(Color online) {\em Time-dependence of Bogoliubov excitation frequencies and density-density correlations.} 
The Bogoliubov excitation energy $\omega_\bk$ (plots (a) and (c)) and the corresponding correlation function in Eq.~\eqref{cf2} (plots (b) and (d)) for zero temperature as a function of the parametrized time $mc^2_f\tau$. Here we fix $k\xi_f\coloneqq 1$, $c^2_i/c^2_f=1/2$. The rate of change in \eqref{c-time} is taken as $a/\omega_{\bk i}=0.3$ for plots (a) and (b), and $a/\omega_{\bk i}=1$ for plots (c) and (d). The black solid curves correspond to contact interaction, $R=0$ ($\tilde{A}=0.1$). The DDI dominated case ($R=\sqrt{\pi/2}$) with varying values of $\tilde{A}$, specified in the insets of (a) and (c), is represented by the other curves.}
\label{fig3}
\end{figure*}

We now impose a time-dependent background by assuming that $c^2=c^2(\tau)$ is of the form 
\begin{eqnarray}\label{c-time}
\frac{c^2(\tau)}{c^2_f}=\frac{1}{2}\bigg(1+\frac{c^2_i}{c^2_f}\bigg)+\frac{1}{2}\bigg(1-\frac{c^2_i}{c^2_f}\bigg)\tanh(a\tau).
\end{eqnarray}
We choose this form of the quench of the sound speed for a direct comparison with the results of 
\cite{PhysRevD.95.065020}, and do indeed find that $R=0$ reproduces the results of the latter reference. 
The above $c^2(\tau)$, in particular, 
implies two asymptotic values, $c^2_i=c^2_0$ and $c^2_f$ which are obtained when $\tau\rightarrow-\infty$ and   $\tau\rightarrow\infty$, respectively, and for which the gas and thus the quasiparticle vacuum become stationary.  
In the examples below, we quench the system to a larger sound speed $c_f>c_i$.

According to \eqref{f-square} and \eqref{speed}, for constant $g_d$ and $g_c$, the scale factor is, given
a prescribed form of $c^2(\tau)$ as in \eqref{c-time}
\begin{equation}\label{scale-factor}
b (\tau)= \frac1{f^2(\tau)} = \frac{c_0^2}{c^2(\tau)}. 
\end{equation} 
The gas, for $c_f>c_i$, therefore contracts, with $b(t_f)<b(t_i)$. 
We plot the scale factor $b(t)$ with respect to the lab time $t$ in Fig.\,\ref{figb}, for various 
quench rates $a$ of the speed of sound in \eqref{c-time}.

As a consequence of the temporal change of $c^2$, 
the quasiparticle state is probed by the operators $\hvphi_{\pm\bk}(\tau)$ whose equation of motion is 
Eq.~\eqref{eqnomega}.
From the time dependence of the excitation frequencies $\omega_\bk(\tau)$, 
the Bogoliubov coefficients $\alpha_\bk(\tau)$ and $\beta_\bk(\tau)$ are 
functions of scaling time $\tau$ as well, satisfying the evolution equations \eqref{eqab}. 
The corresponding correlation function in the first line of Eq.~\eqref{cf1} becomes 
\begin{eqnarray}\label{cf2}
\nonumber
G_{2, \bk}(\tau)&=&(u_{\bk}(\tau)+v_{\bk}(\tau))^2 \big[|\alpha_{\bk}(\tau)|^2+|\beta_{\bk}(\tau)|^2
\\   \nonumber
&&+2\Re\big\{\alpha_{\bk}(\tau)\beta^\ast_{\bk}(\tau)e^{-2i\omega_{\bk}\tau^\prime}\big\}\big] (2n^\mathrm{in}_{\bk}+1).
\\
\end{eqnarray}
We can rewrite Eq.~\eqref{cf2} in the form of Eq.~\eqref{cf1}, with 
\begin{eqnarray}\label{final-nk}
\nonumber
2n_\bk+1&=&(|\alpha_\bk|^2+|\beta_\bk|^2)(2n^\mathrm{in}_\bk+1),
\\
c_\bk&=&\alpha_\bk\beta^\ast_\bk(2n^\mathrm{in}_\bk+1).
\end{eqnarray}

For adiabatic variations ($a\rightarrow 0$ in \eqref{c-time}), 
$\alpha_\bk$ and $\beta_\bk$ do essentially not change and remain very close to $1$ and $0$, respectively; $G_{2, \bk}$ then varies in time only because $(u_\bk+v_\bk)^2$ does so. 
However, when we are in a nonadiabatic regime, $\alpha_\bk$ and $\beta_\bk$ evolve in time, and 
the degree of nonadiabaticity is encoded in them. 
We conclude from Eq.~\eqref{final-nk} that initial thermal quasiparticle noise can enhance
quasiparticle production because the quantities shown in \eqref{final-nk}, which occur
in \eqref{cf1}, are proportional to the initial thermal background multiplicative factor $2n^\mathrm{in}_\bk+1$.
{However, let us note that while the quasiparticle production is enhanced as regards their sheer number, this happens at the expense of entanglement, since the thermally stimulated pairs are not quantum mechanically correlated.}

With the time evolution of $c^2(\tau)$ as prescribed in Eq.~\eqref{c-time}, we obtain the time-dependent   
$\omega_\bk(\tau)$ in \eqref{omega1}, and then 
can solve the  coupled Eqs.~\eqref{eqab} numerically. In Fig.\,\ref{fig3} we show some examples of the evolution of the quasiparticle frequencies at fixed momentum (left panel), and the corresponding correlation function response in Eq.~\eqref{cf2} to this evolution (right panel). In what follows, we will use the following definitions of {healing length} and effective chemical potential, respectively:
\begin{eqnarray}
\xi_f= \frac1{mc_f},~~~~~\tilde{A}=\frac{mc^2_f}{\omega_{z,0}}=A\frac{c^2_f}{c^2_i}. \label{def_tildeA}
\end{eqnarray}
{Note that $\omega_{z,0}$ is constant in the comoving frame. Hence $\tilde{A}/A=c^2_f/c^2_i$
depends on the initial and final speeds of sound only.}

The quasiparticle frequencies approach two asymptotics 
because $c^2(\tau)$ approaches constants 
in the limits of $\tau\rightarrow-\infty$ and $\tau\rightarrow\infty$, respectively (left panel of Fig.\,\ref{fig3}).  
For $k\xi_f=1$, when $\tilde{A}<1.073$, the initial frequencies 
$\omega_{\bk\,i}=\lim_{\tau\rightarrow-\infty}\omega_\bk(\tau)$ are smaller than the final 
frequencies $\omega_{\bk\,f}=\lim_{\tau\rightarrow\infty}\omega_\bk(\tau)$. However, when 
$\tilde{A}$ is large (assuming DDI {dominance}, $R=\sqrt{\pi/2}$), i.e., $1.073<\tilde{A}\le3.4454$ for $k\xi_f=1$, 
the initial frequencies 
$\omega_{\bk\,i}=\lim_{\tau\rightarrow-\infty}\omega_\bk(\tau)$ are larger than the corresponding final ones, $\omega_{\bk\,f}=\lim_{\tau\rightarrow\infty}\omega_\bk(\tau)$.  
{Therefore (a dominant) DDI, together with a sufficiently high (but experimentally feasible, cf.~ the discussion in \cite{PhysRevA.73.031602}) density of the gas, which are parametrized by $R$ and  $\tilde A$,  respectively, can significantly affect the quasiparticle response when comparing with a gas possessing only contact interactions.}
In the presence of a (sufficiently strong) DDI, a roton minimum appears. For increasing roton depth,  finite-momentum excitation frequencies near the roton minimum are small; hence these modes are more sensitive to temporal changes of the background.

In the two asymptotical regimes, one has a well defined vacuum for the quasiparticles. These vacua are not necessarily equivalent to each other. 
The vacuum defined in the far-past region are seen as a two-mode squeezed state from the viewpoint of the observer in the far-future region.
That is to say, although there are no quasiparticles at the beginning, due to an 
expansion or contraction of the condensate, excitations will be created from the quasiparticle vacuum. 

The temporal behavior of the correlation function in Eq.~\eqref{cf2} is strongly affected by the 
strength of the DDI and the gas density (see right panel of Fig.\,\ref{fig3}).
When the variation in time of $\omega_\bk$ is slow, i.e., when $a$ is small, 
$G_{2, \bk}(\tau)$ varies smoothly for small $\tilde{A}$ ($\tilde{A}<1.073$ in Fig.\,\ref{fig3}). 
When the change of $c^2$ is sufficiently abrupt, the two-point density correlation function 
oscillates such that it can periodically dip below its vacuum value. 
For large $\tilde{A}$ ($\tilde{A}>1.073$ in Fig.\,\ref{fig3}), 
the corresponding two-point density correlations oscillate with larger amplitude 
than for smaller $\tilde{A}$ (smaller chemical potential).

{It is interesting to note that when $k\xi_f=1$ and $\tilde{A}\approx1.073$ and with DDI dominating  
(i.e., $R=\sqrt{\pi/2}$), the {frequency} of quasiparticles is time-independent. The reason is that in this case the interaction energy, $\mathcal{A}_{\mathbf{k}}$, shown in Eq. \eqref{Ak_def} is equal to zero, and thus the {frequency} given in Eq. \eqref{omega1} is reduced to the single-particle part $\mathcal{H}_\mathbf{k}$,  which is time-independent. This, then, implies that there are no quasiparticles created by the quench. We thus find that the corresponding density-density correlations are time-independent as well.}

\begin{figure*}[p]
\centering
\subfigure[~~\normalsize{$T=0$}]{\includegraphics[width=0.36\textwidth]{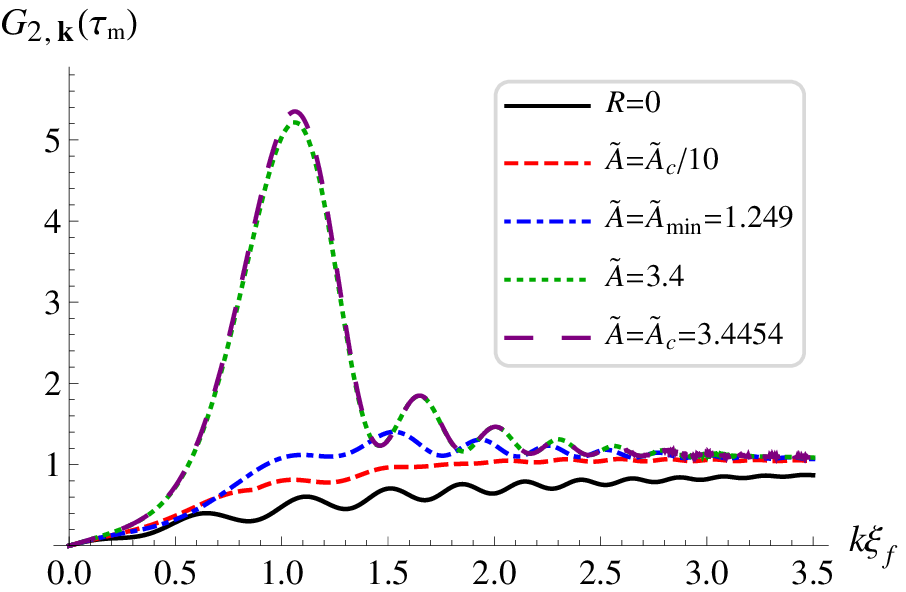}}\hspace*{2em}
\subfigure[~~\normalsize{$T=mc^2_0$}]{\includegraphics[width=0.36\textwidth]{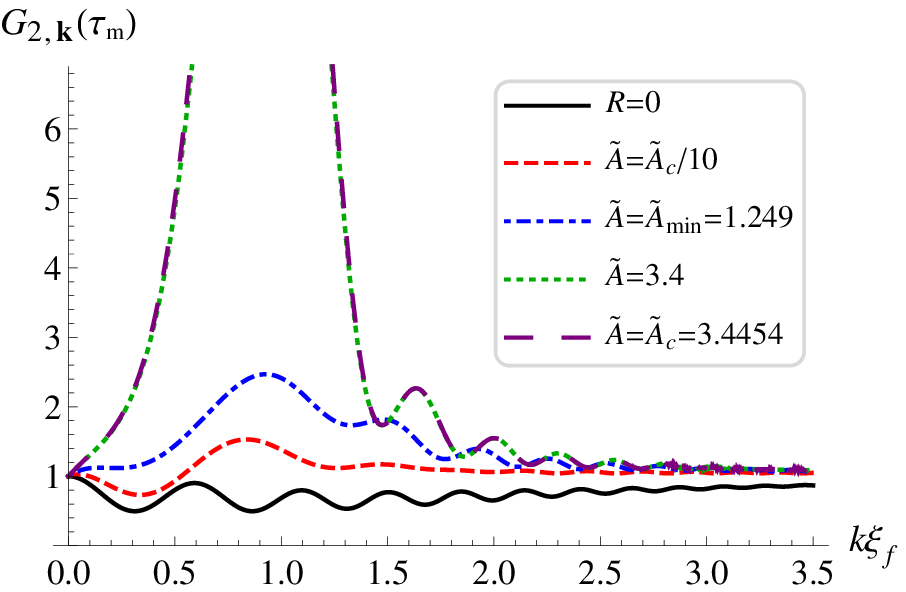}}
\subfigure[~~\normalsize{$T=0$}]{\includegraphics[width=0.36\textwidth]{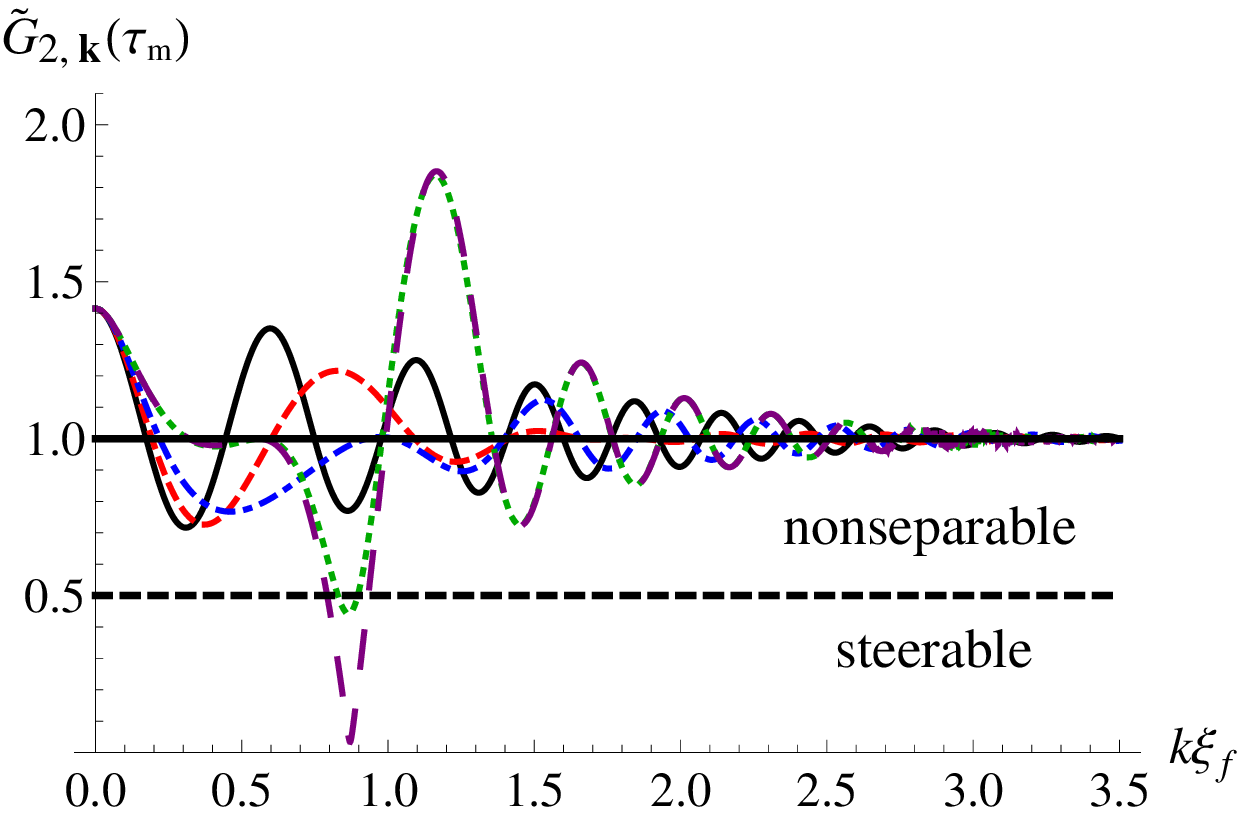}}\hspace*{2em}
\subfigure[~~\normalsize{$T=mc^2_0$}]{\includegraphics[width=0.36\textwidth]{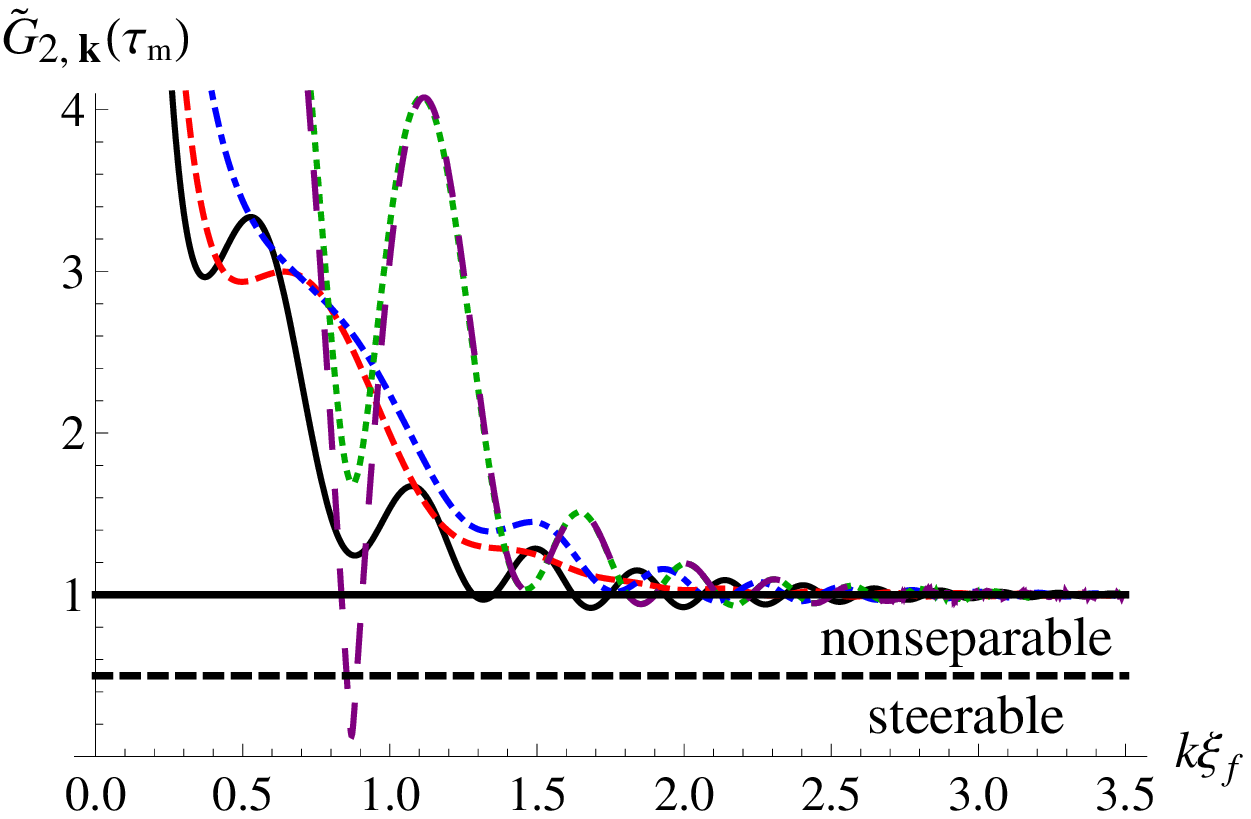}}
\caption{(Color online) 
{\em Density-density correlations as a function of $k\xi_f$ at zero temperature (left) and finite temperature (right).}
The measurement time is $\tau_{\rm m}=5\times (mc^2_f)^{-1}$. 
Here $c^2_i/c^2_f=1/2$, and the rate of change $a/\omega_{\bk i}=1$($k\xi_f=3$). The solid curve corresponds to contact interaction, $R=0$ ($\tilde{A}=\tilde{A}_c/10$). 
DDI {dominance} ($R=\sqrt{\pi/2}$) for the other curves, with 
$\tilde{A}$ 
specified in the insets of (a) and (b). 
The lower plots show correlation functions normalized by $(u_\bk+v_\bk)^2$, such that the nonseparability and steerability thresholds occur at $1$ (thick black line) and $1/2$ (dashed thick black line), respectively.
\label{Fig4}}
\centering
\subfigure[~~\normalsize{$T=0$}]{\includegraphics[width=0.36\textwidth]{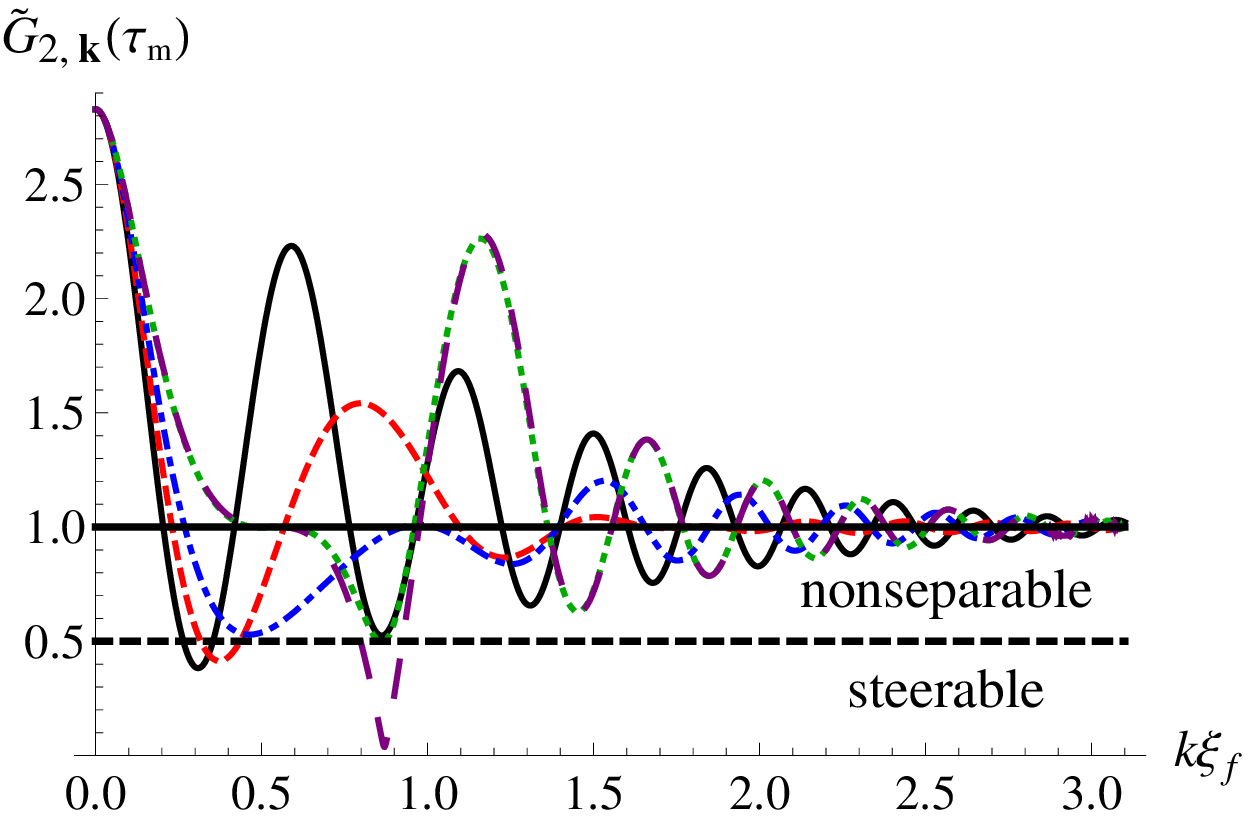}}\hspace*{4em}
\subfigure[~~\normalsize{$T=mc^2_0$}]{\includegraphics[width=0.36\textwidth]{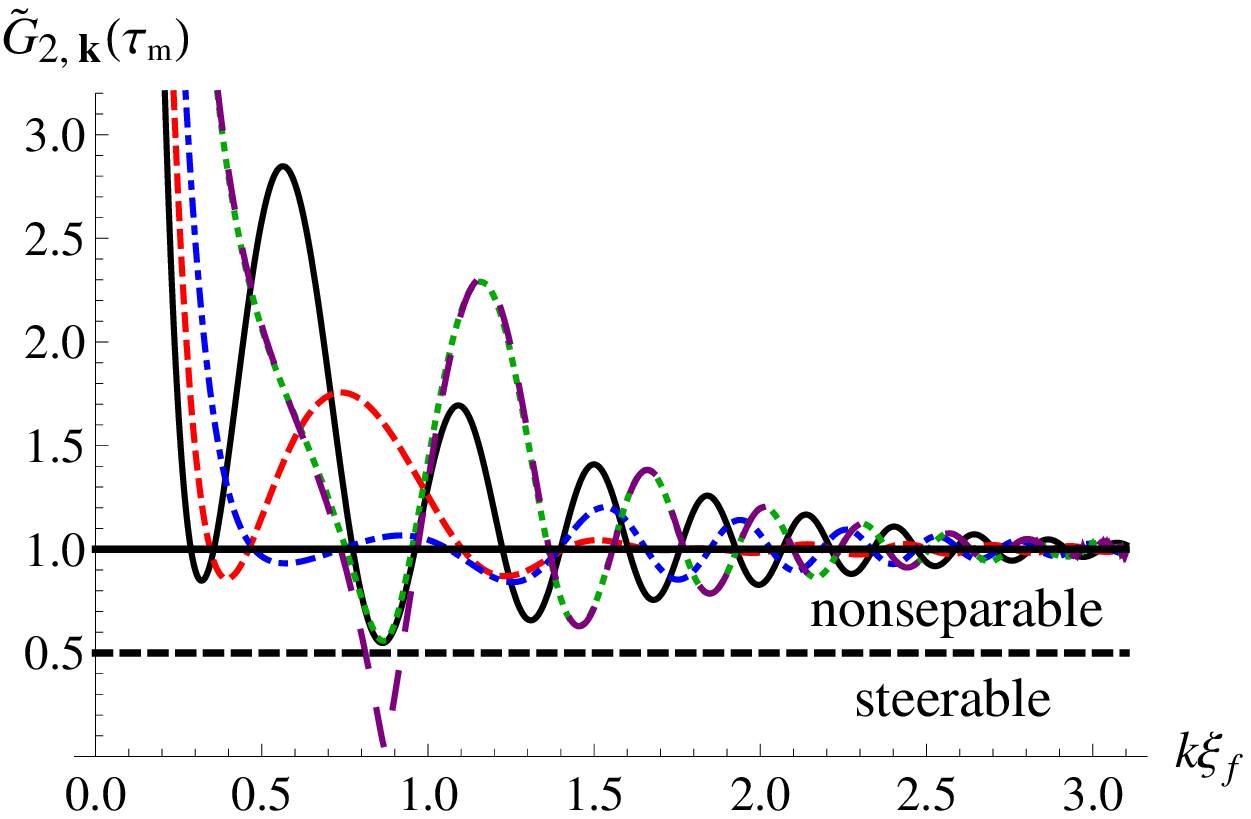}}
\subfigure[~~\normalsize{$T=0$}]{\includegraphics[width=0.36\textwidth]{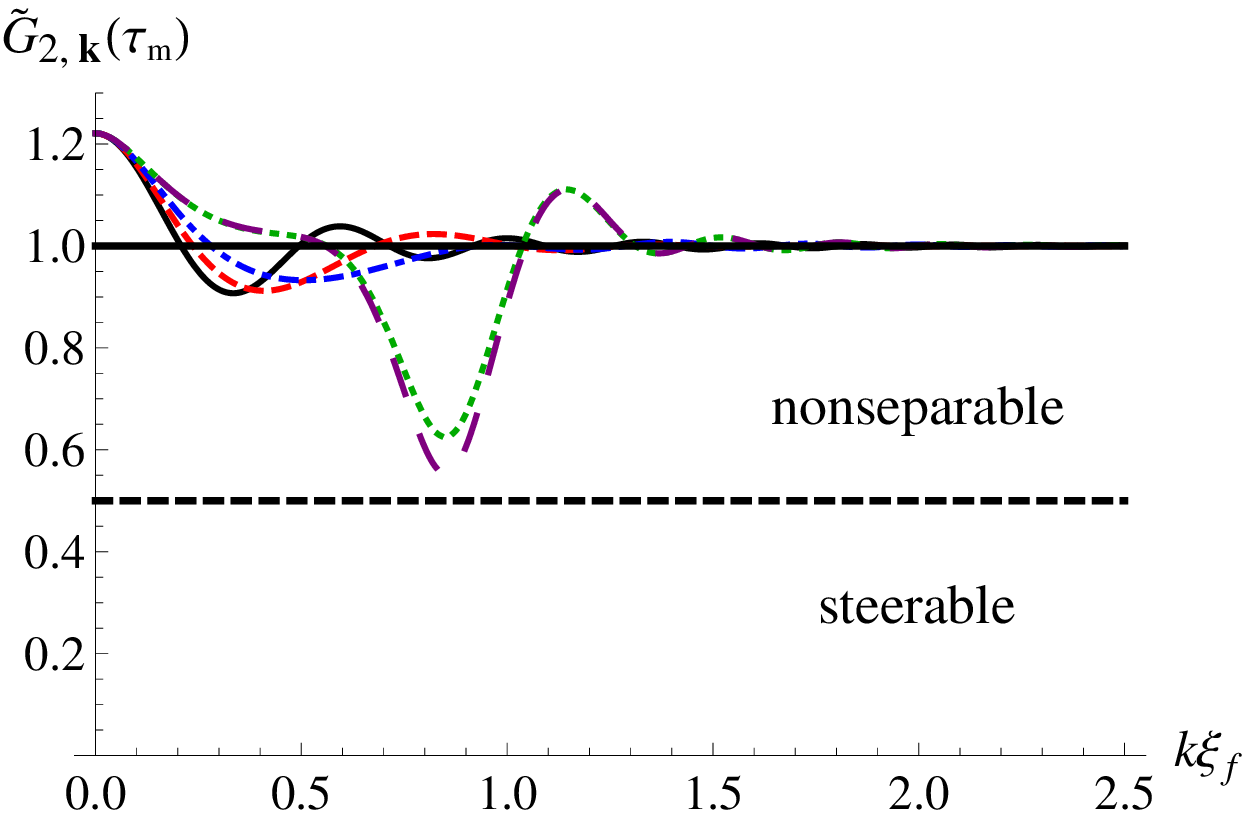}}\hspace*{4em}
\subfigure[~~\normalsize{$T=mc^2_0$}]{\includegraphics[width=0.36\textwidth]{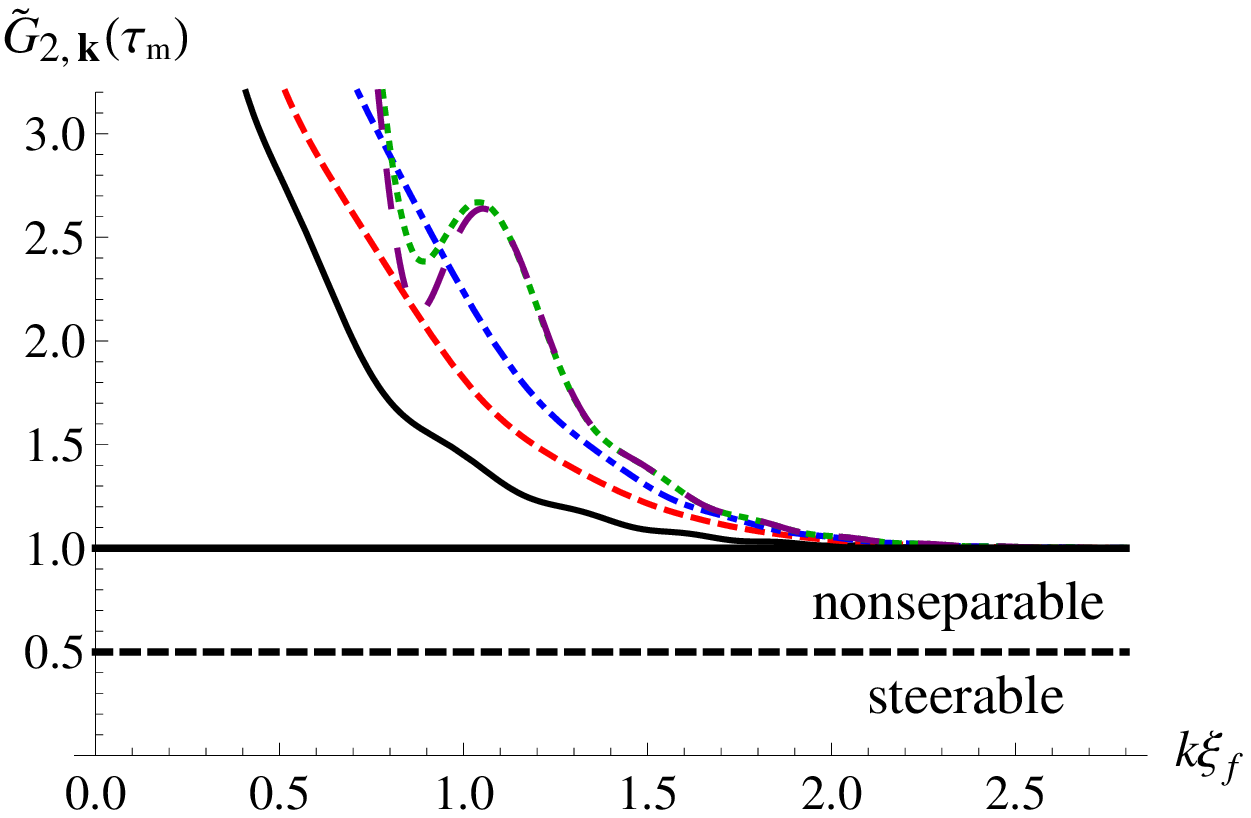}}
\caption{(Color online) {\em Varying 
$c^2_f$ and sweep rate $a$  for zero temperature (left) and finite temperature (right).} 
Shown is the variation of the 
normalized density-density correlation functions with $k\xi_f$ at fixed measurement 
time $\tau_{\rm m}=5\times (mc^2_f)^{-1}$. 
{\em (a) and (b)} Larger final sound speed $c^2_i/c^2_f=1/8$ than in Fig.\,\ref{Fig4} {(}a) and (b), with  identical rate of change $a/\omega_{\bk i}=1$ ($k\xi_f=3$). 
{\em (c) and (d)} Smaller sweep rate than in Fig.\,\ref{Fig4} {(}c) and (d), with  identical
$c^2_i/c^2_f=1/2$, and the rate of change $a/\omega_{\bk i}=0.05$ ($k\xi_f=3$). 
The values of $\tilde{A}$ corresponding to the various curves are 
found in the insets of Fig.\,\ref{Fig4}~(a) and (b). 
\label{Fig5}}
\end{figure*}

\subsection{Quench production of nonseparability and {steerability}} 

\begin{figure*}[t]
\centering
\subfigure[~~\normalsize{$T=0$}]{\includegraphics[width=0.36\textwidth]{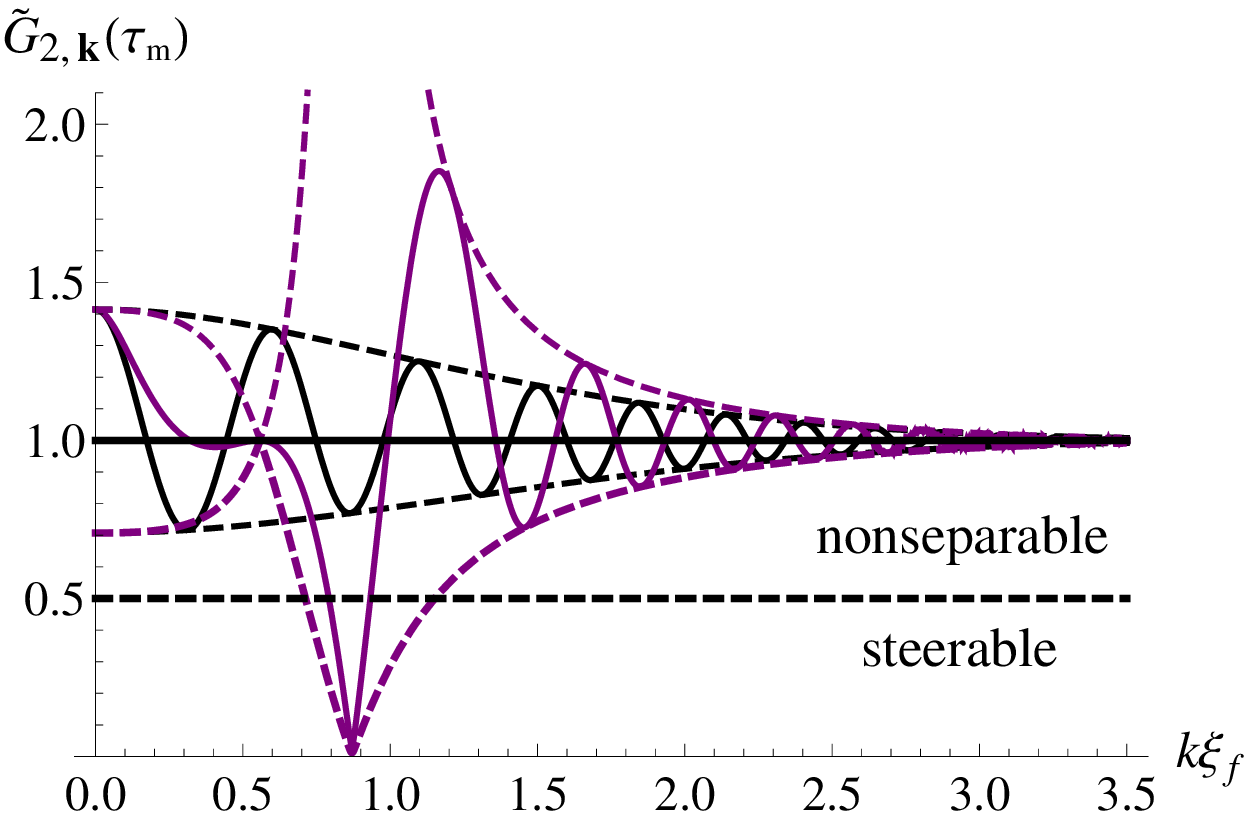}}\hspace*{2em}
\subfigure[~~\normalsize{$T=mc^2_0$}]{\includegraphics[width=0.36\textwidth]{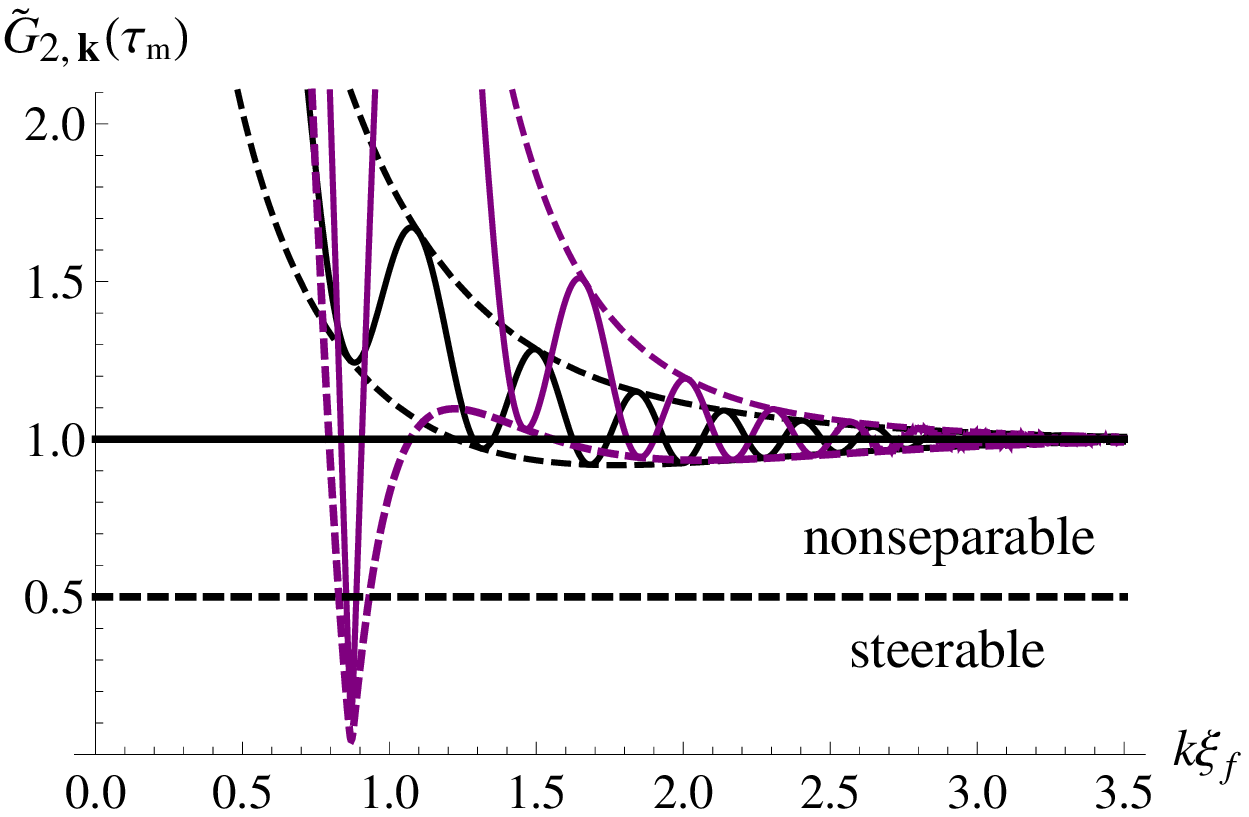}}
\subfigure[~~\normalsize{$T=0$}]{\includegraphics[width=0.36\textwidth]{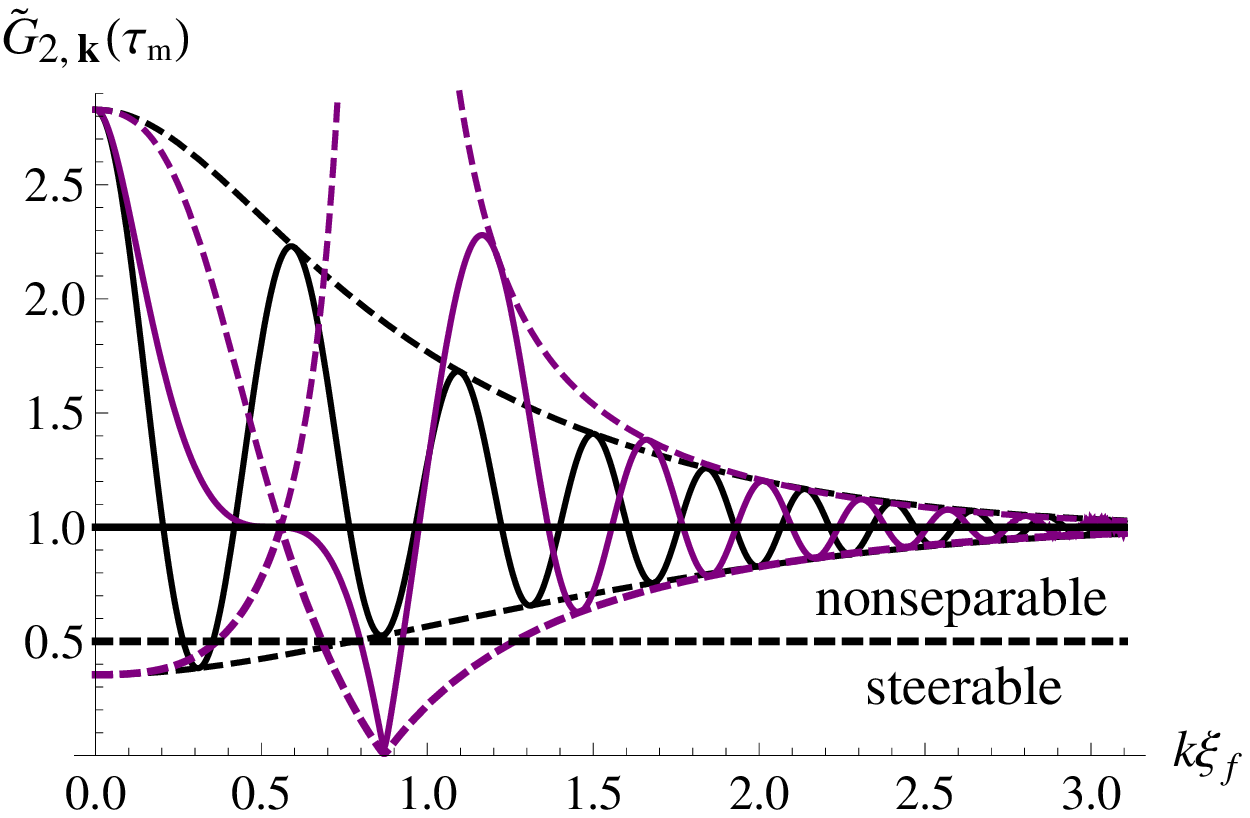}}\hspace*{2em}
\subfigure[~~\normalsize{$T=mc^2_0$}]{\includegraphics[width=0.36\textwidth]{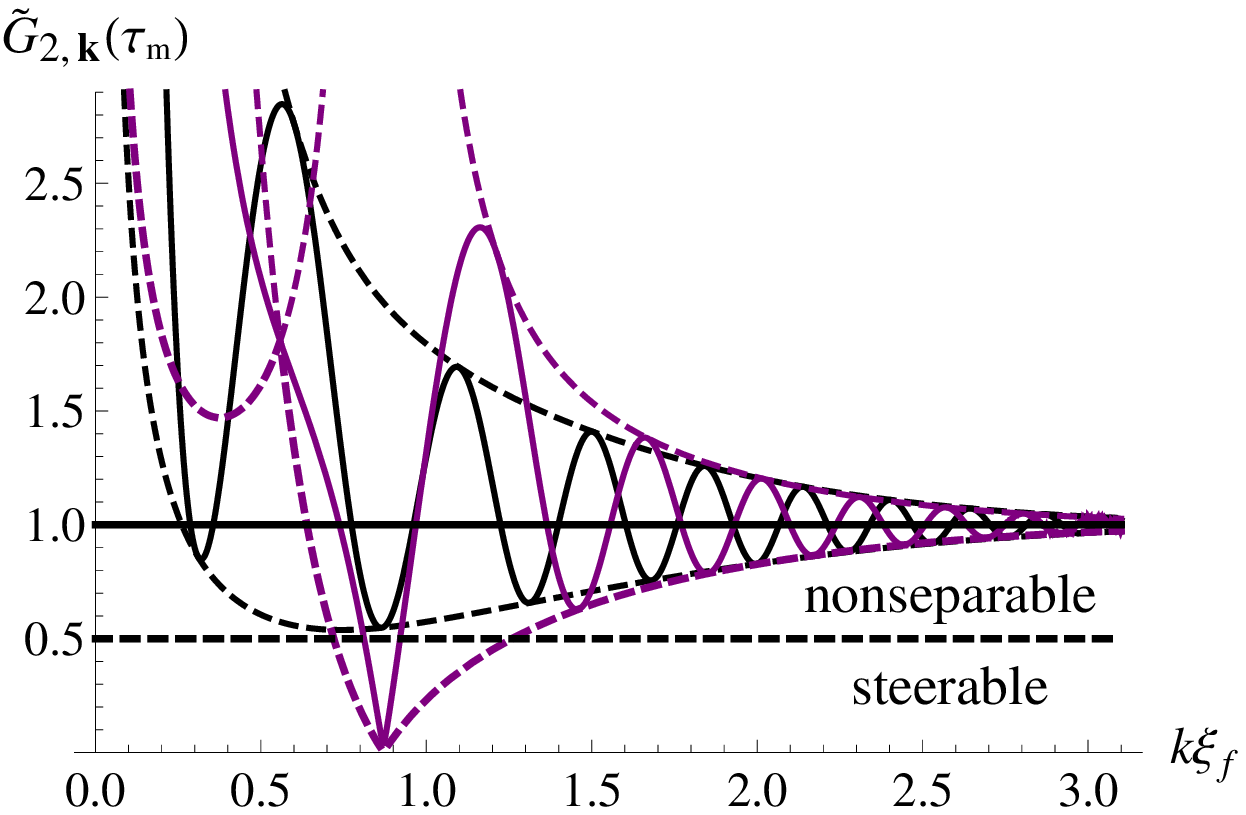}}
\subfigure[~~\normalsize{$T=0$}]{\includegraphics[width=0.36\textwidth]{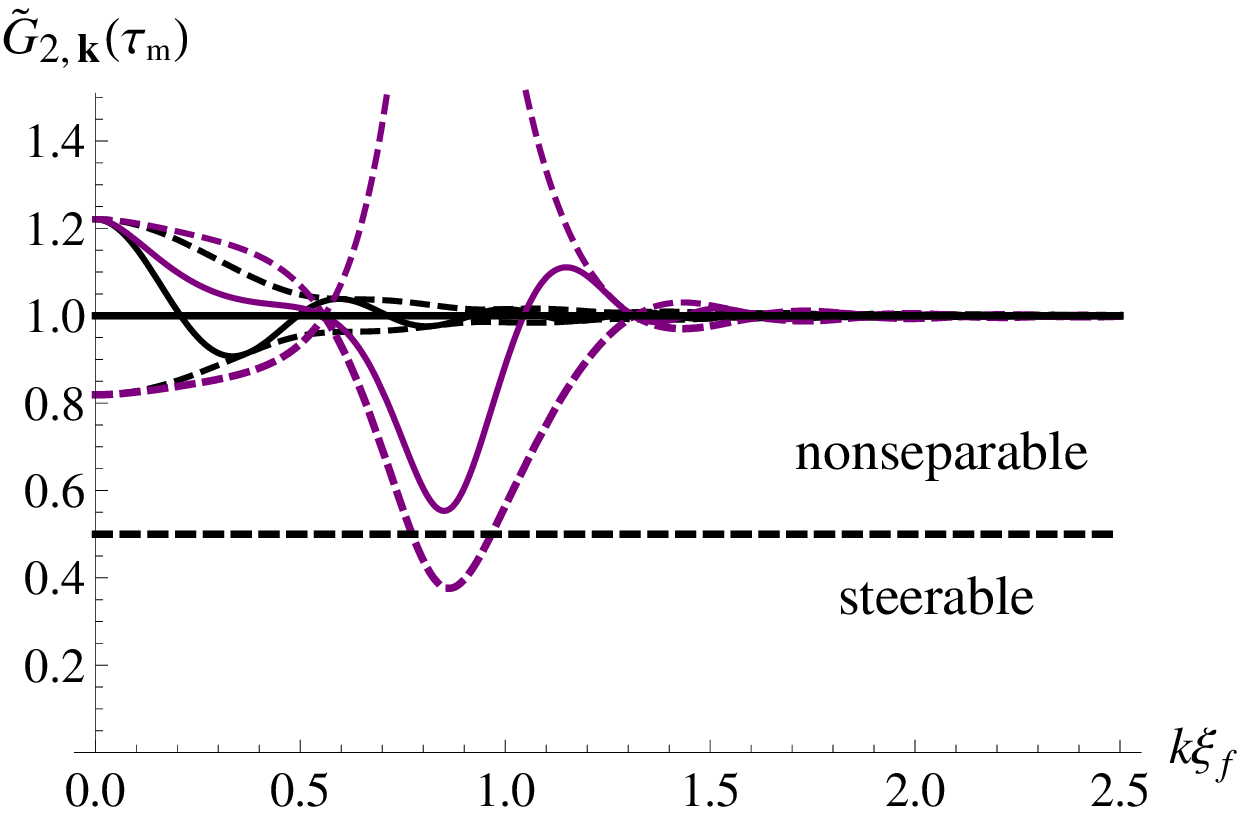}}\hspace*{2em}
\subfigure[~~\normalsize{$T=mc^2_0$}]{\includegraphics[width=0.36\textwidth]{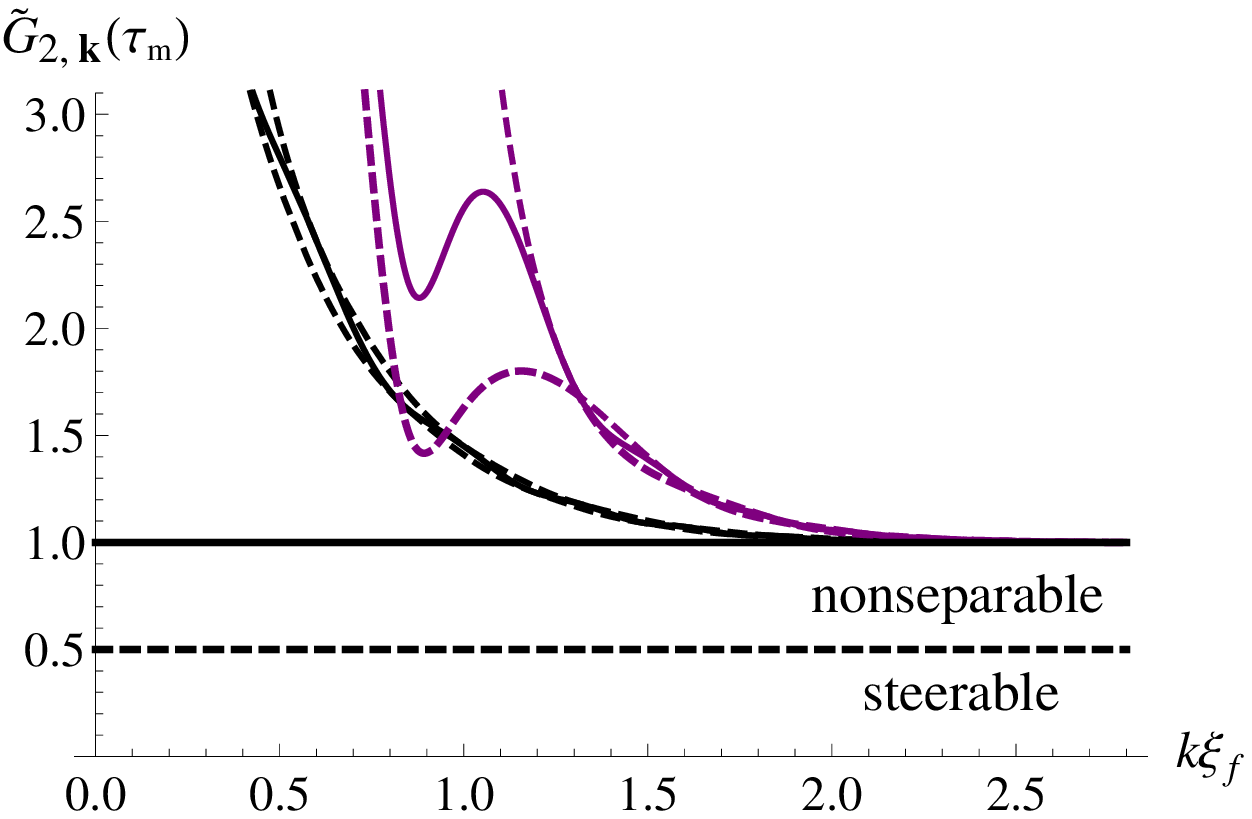}}
\caption{(Color online) 
{{\em Density-density correlations as a function of $k\xi_f$ at zero temperature (left) and finite temperature (right).}
The measurement time is $\tau_{\rm m}=5\times (mc^2_f)^{-1}$. 
Here $c^2_i/c^2_f=1/2$, and the rate of change $a/\omega_{\bk i}=1$ ($k\xi_f=3$) for (a) and (b); $c^2_i/c^2_f=1/8$, and the rate of change $a/\omega_{\bk i}=1$ ($k\xi_f=3$) for (c) and (d). Finally, $c^2_i/c^2_f=1/2$, and the rate of change $a/\omega_{\bk i}=0.05$ ($k\xi_f=3$) for (e) and (f). The black solid curves corresponds to contact interaction, $R=0$ ($\tilde{A}=\tilde{A}_c/10$). 
The solid purple curves are for the DDI-dominated case ($R=\sqrt{\pi/2}$) at criticality, $\tilde{A}=3.4454$. 
The dashed lines are envelopes.
Correlation functions are normalized by $(u_\bk+v_\bk)^2$, such that the nonseparability and steerability thresholds occur at $1$ (thick black line) and $1/2$ (dashed thick black line), respectively.
\label{Fig6}}}
\end{figure*}

In an experiment, a measurement is performed on the condensate at some given 
time $\tau_{\rm m}$. 
To study quantum correlations between the produced quasiparticle modes,
we focus here on the variation of the correlation function with momentum $k\xi_f$ at fixed time 
$\tau_{\rm m}$. 
As an example, we plot in Fig.\,\ref{Fig4}
the variation of the correlation function in the momentum at fixed measurement time $\tau=\tau_{\rm m}$.
To examine nonseparability and {steerability} between the produced quasiparticles, we plot the normalized correlation function, i.e. the correlation function divided by its vacuum value 
\begin{equation}\label{NCF}
{\tilde G}_{2,\bk}(\tau) \coloneqq \frac{G_{2,\bk}}{G_{2, \bk}^\mathrm{vac}} =  \frac{G_{2,\bk}}{(u_\bk+v_\bk)^2} .
\end{equation}
The nonseparability and steerability thresholds  then occur according to Eqs.~\eqref{ie2} and  \eqref{QS3}
at ${\tilde G}_{2,\bk} =1$ and ${\tilde G}_{2,\bk} =1/2$, respectively. 

{The normalized density-density correlation function periodically changes and potentially dips below unity. 
When the normalized density-density correlation function is smaller than $1$ for some times, the final quasiparticle state is nonseparable.}
This implies that entanglement is created between quasiparticles with opposite momenta $\bk$ and $-\bk$ due to the nonadiabatic variation of the speed of sound of the condensate and the excitation of the condensate vacuum. Moreover, even though 
initial thermal noise  decreases the range of nonseparable $\bk$'s (right panel of Fig.\,\ref{Fig4}), a sufficiently dense dipolar gas close to criticality still creates entanglement (comparing (d) with (c) in Fig.\,\ref{Fig4}). {In particular, looking at Fig. 5(d),} the momentum which renders the final quasiparticle state nonseparable, that is which satisfies the inequality \eqref{ie2}, is for the dipolar gas  
smaller than for a contact interaction gas. 

{Although the sufficient condition \eqref{QS3} for steerability might not be
satisfied for any value of $\bk$ in the final quasiparticle state when
only contact interactions are present, a sufficiently large DDI rather generically induces a
state which does satisfy this criterion {for some  momenta} 
 (see the green dotted and purple long-dashed curves in (c) of Fig.\,\ref{Fig4}).} 
As discussed in subsection \ref{steersection},  
steerability is a more stringent correlation property of quantum states 
than nonseparability is (however, still weaker than Bell nonlocality). 
Steering implies that the state is nonseparable but not vice versa, a fact 
which is readily confirmed with Figs.\,\ref{Fig4} and \ref{Fig5}. 

The time-dependent speed of sound as specified in \eqref{c-time} can, for example, 
be adjusted by the external potential trapping the condensate, 
according to the scaling equations \eqref{ST} and \eqref{f-square}. To determine  
how the quench rate and final sound speed squared, $c^2_f$, affect the creation of quasiparticle entanglement, we plot the normalized density-density correlation functions in Fig.\,\ref{Fig5}. 
We conclude that quantum steering between quasiparticles is robustly obtained whenever we are near criticality $\tilde{A}\lesssim \tilde{A}_\mathrm{c}$. In addition, we observe that 
an increase of $c_f^2$ amplifies the fluctuations of the normalized density-density correlation functions around their mean values (comparing (a) in Fig.\,\ref{Fig5} with (c) in Fig.\,\ref{Fig4}), and induces the creation of quasiparticle steering in a condensate with relatively low density ($\tilde{A}<\tilde{A}_\mathrm{min}$). 
On the other hand,  smaller sweep rates $a/\omega_{\bk i}$ decrease the amplitude of the fluctuations of the normalized density-density correlation functions (comparing (c) in Fig.\,\ref{Fig5} with (c) in Fig.\,\ref{Fig4}); {they also decrease the production of quasiparticle steering near criticality, especially for the thermal case (more details are revealed from inspecting {t}he envelopes in Fig. \ref{Fig6}).}  

{{After the quench, the final spectrum $\omega_{\bk}$ of quasiparticles becomes time-independent, and the factor $e^{-2i\int^\tau\omega_{\bk}(\tau^\prime)d\tau^\prime}$ 
in Eq. \eqref{cf2} is then simply $e^{-2i\omega_{\bk}\tau}$.}  
The density-density correlation function \eqref{cf2} or its normalized version \eqref{NCF} then periodically oscillates. We therefore can verify whether the final quasiparticles are entangled by looking at the corresponding  envelopes, i.e., by determining the maximum and minimum values reached by the  density-density correlation function \eqref{NCF} as it oscillates in time. With the criteria for nonseparability and steerability displayed in \eqref{ie2} and \eqref{QS3}, respectively, the quasiparticles are  
entangled or even steerable when the lower envelope for the normalized correlations dips below $1$ and $1/2$, respectively.}

To show the domains of nonseparability and steerability more clearly, and make the comparison between the contact interaction case and the DDI case more readily accessible, in Fig.\,\ref{Fig6} we plot the envelopes for the contact interaction case and the DDI case with the critical value of $\tilde{A}=3.4454$ shown in Figs.\,\ref{Fig4} and \ref{Fig5}. 
{We conclude that, when DDI dominates, the created quasiparticles with wave vectors near the roton minimum satisfy the sufficient criterion \eqref{QS3} for steerability, while this is not the case for contact interactions.} Therefore, we conclude that compared to a gas with solely repulsive contact interactions, the DDI Bose gas system displays an enhanced potential for the presence of steering in the bipartite quantum state of quasiparticle pairs resulting from the quench. {This enhancement is generally robust against thermal noise, and variation of the difference between the initial and final speeds of sound as well as of the quench rate. }

{One notices from Fig.~\ref{Fig6} that there is an exceptional point where the upper and the lower envelopes cross. At this point, {the normalized density-density correlation functions 
are equal to unity in the zero temperature limit}.  {This requires that the Bogoliubov coefficient $\beta_\bk(\tau)$ vanishes, implying that for these momenta there are no quasiparticles created. }
This, in turn, is due to the fact that the interaction energy ${\mathcal A}_\bk(\tau)$ in \eqref{omega1} equals zero, 
and thus the spectrum is time-independent. Vanishing  ${\mathcal A}_\bk(\tau)$ is 
due to the fact that the Fourier transform of the interaction $V_{\intac,0}^{\td}(k)$  
in \eqref{Ak_def} crosses zero for a certain momentum when $R>\frac23 \sqrt{\pi/2}$ (i.e., when 
$g_{d,0}>g_{c,0}$ \cite{PhysRevA.73.031602}). The phenomenon just described  
does not happen for contact interactions, where the Fourier transform is simply a constant, {so that $\beta_\bk$ is non-zero for all momenta. }

{Let us, finally, note from Fig.~\ref{Fig6} 
that the maxima of the upper envelopes at the roton minimum are very large, and that the momentum range for the lower envelopes to satisfy the steerability criterion is relatively narrow. Hence, care in choosing the appropriate measurement time $\tau_{\rm m}$ and appropriate momentum $\bk$ needs to be exercised, to reliably judge whether in a given experiment the created quasiparticles 
satisfy the entanglement criteria.}}

\section{Conclusion} \label{section4}
We have studied the production of quasiparticle pairs in a quasi-two-dimensional dipolar condensate undergoing a  rapid temporal variation of its speed of sound, 
and focused on criteria using the experimentally accessible density-density correlation function \cite{PhysRevD.92.024043}, 
to determine the nonseparability and steerability of the final quasiparticle state. As demonstrated in Figs.\,\ref{Fig4}, \ref{Fig5} and \ref{Fig6},
the DDI between the gas particles significantly enhances the potential for the creation of entanglement 
{(nonseparability and steering)}, being established 
between quasiparticle modes $\bk$ and $-\bk$. 
We conclude that the dipolar two-body interaction leads to increased {quantum correlations}, as encoded in the bipartite continuous-variable state of quasiparticles created by a quench.

\begin{acknowledgments}
ZT and SYC were supported by the BK21 Plus Program (21A20131111123) funded by the Ministry of Education (MOE, Korea) and National Research Foundation of Korea (NRF). 
SYC and URF were supported by the NRF, Grant No.  2017R1A2A2A05001422.

\end{acknowledgments}

\bibliography{steer16}

\begin{thebibliography}{64}%
\makeatletter
\providecommand \@ifxundefined [1]{%
 \@ifx{#1\undefined}
}%
\providecommand \@ifnum [1]{%
 \ifnum #1\expandafter \@firstoftwo
 \else \expandafter \@secondoftwo
 \fi
}%
\providecommand \@ifx [1]{%
 \ifx #1\expandafter \@firstoftwo
 \else \expandafter \@secondoftwo
 \fi
}%
\providecommand \natexlab [1]{#1}%
\providecommand \enquote  [1]{``#1''}%
\providecommand \bibnamefont  [1]{#1}%
\providecommand \bibfnamefont [1]{#1}%
\providecommand \citenamefont [1]{#1}%
\providecommand \href@noop [0]{\@secondoftwo}%
\providecommand \href [0]{\begingroup \@sanitize@url \@href}%
\providecommand \@href[1]{\@@startlink{#1}\@@href}%
\providecommand \@@href[1]{\endgroup#1\@@endlink}%
\providecommand \@sanitize@url [0]{\catcode `\\12\catcode `\$12\catcode
  `\&12\catcode `\#12\catcode `\^12\catcode `\_12\catcode `\%12\relax}%
\providecommand \@@startlink[1]{}%
\providecommand \@@endlink[0]{}%
\providecommand \url  [0]{\begingroup\@sanitize@url \@url }%
\providecommand \@url [1]{\endgroup\@href {#1}{\urlprefix }}%
\providecommand \urlprefix  [0]{URL }%
\providecommand \Eprint [0]{\href }%
\providecommand \doibase [0]{http://dx.doi.org/}%
\providecommand \selectlanguage [0]{\@gobble}%
\providecommand \bibinfo  [0]{\@secondoftwo}%
\providecommand \bibfield  [0]{\@secondoftwo}%
\providecommand \translation [1]{[#1]}%
\providecommand \BibitemOpen [0]{}%
\providecommand \bibitemStop [0]{}%
\providecommand \bibitemNoStop [0]{.\EOS\space}%
\providecommand \EOS [0]{\spacefactor3000\relax}%
\providecommand \BibitemShut  [1]{\csname bibitem#1\endcsname}%
\let\auto@bib@innerbib\@empty
\bibitem [{\citenamefont {Birrell}\ and\ \citenamefont
  {Davies}(1982)}]{Quantumfield}%
  \BibitemOpen
  \bibfield  {author} {\bibinfo {author} {\bibfnamefont {N.~D.}\ \bibnamefont
  {Birrell}}\ and\ \bibinfo {author} {\bibfnamefont {P.~C.~W.}\ \bibnamefont
  {Davies}},\ }\enquote {\bibinfo {title} {{{Quantum Fields in Curved
  Space}}},}\ \ (\bibinfo  {publisher} {Cambridge University Press, Cambridge,
  England},\ \bibinfo {year} {1982})\BibitemShut {NoStop}%
\bibitem [{\citenamefont {Schr\"odinger}(1939)}]{Schroedinger}%
  \BibitemOpen
  \bibfield  {author} {\bibinfo {author} {\bibfnamefont {Erwin}\ \bibnamefont
  {Schr\"odinger}},\ }\bibfield  {title} {\enquote {\bibinfo {title} {The
  proper vibrations of the expanding universe},}\ }\href {\doibase
  https://doi.org/10.1016/S0031-8914(39)90091-1} {\bibfield  {journal}
  {\bibinfo  {journal} {Physica}\ }\textbf {\bibinfo {volume} {6}},\ \bibinfo
  {pages} {899--912} (\bibinfo {year} {1939})}\BibitemShut {NoStop}%
\bibitem [{\citenamefont {Parker}(1968)}]{PhysRevLett.21.562}%
  \BibitemOpen
  \bibfield  {author} {\bibinfo {author} {\bibfnamefont {L.}~\bibnamefont
  {Parker}},\ }\bibfield  {title} {\enquote {\bibinfo {title} {{Particle
  Creation in Expanding Universes}},}\ }\href {\doibase
  10.1103/PhysRevLett.21.562} {\bibfield  {journal} {\bibinfo  {journal} {Phys.
  Rev. Lett.}\ }\textbf {\bibinfo {volume} {21}},\ \bibinfo {pages} {562--564}
  (\bibinfo {year} {1968})}\BibitemShut {NoStop}%
\bibitem [{\citenamefont {Moore}(1970)}]{DynCasimir}%
  \BibitemOpen
  \bibfield  {author} {\bibinfo {author} {\bibfnamefont {Gerald~T.}\
  \bibnamefont {Moore}},\ }\bibfield  {title} {\enquote {\bibinfo {title}
  {{Quantum Theory of the Electromagnetic Field in a Variable-Length
  One-Dimensional Cavity}},}\ }\href {\doibase 10.1063/1.1665432} {\bibfield
  {journal} {\bibinfo  {journal} {Journal of Mathematical Physics}\ }\textbf
  {\bibinfo {volume} {11}},\ \bibinfo {pages} {2679--2691} (\bibinfo {year}
  {1970})}\BibitemShut {NoStop}%
\bibitem [{\citenamefont {Hawking}(1975)}]{Hawking1975}%
  \BibitemOpen
  \bibfield  {author} {\bibinfo {author} {\bibfnamefont {S.~W.}\ \bibnamefont
  {Hawking}},\ }\bibfield  {title} {\enquote {\bibinfo {title} {Particle
  creation by black holes},}\ }\href {\doibase 10.1007/BF02345020} {\bibfield
  {journal} {\bibinfo  {journal} {Communications in Mathematical Physics}\
  }\textbf {\bibinfo {volume} {43}},\ \bibinfo {pages} {199--220} (\bibinfo
  {year} {1975})}\BibitemShut {NoStop}%
\bibitem [{\citenamefont {Visser}(1998{\natexlab{a}})}]{Visser1998}%
  \BibitemOpen
  \bibfield  {author} {\bibinfo {author} {\bibfnamefont {M.}~\bibnamefont
  {Visser}},\ }\bibfield  {title} {\enquote {\bibinfo {title} {{Hawking
  Radiation without Black Hole Entropy}},}\ }\href {\doibase
  10.1103/PhysRevLett.80.3436} {\bibfield  {journal} {\bibinfo  {journal}
  {Phys. Rev. Lett.}\ }\textbf {\bibinfo {volume} {80}},\ \bibinfo {pages}
  {3436--3439} (\bibinfo {year} {1998}{\natexlab{a}})}\BibitemShut {NoStop}%
\bibitem [{\citenamefont {Insight}(April 2012)}]{QS}%
  \BibitemOpen
  \bibfield  {author} {\bibinfo {author} {\bibfnamefont {Nature~Physics}\
  \bibnamefont {Insight}},\ }\bibfield  {title} {\enquote {\bibinfo {title}
  {Quantum simulation},}\ }\href {http://dx.doi.org/10.1038/nphys2258}
  {\bibfield  {journal} {\bibinfo  {journal} {Nat. Phys.}\ }\textbf {\bibinfo
  {volume} {8}} (\bibinfo {year} {April 2012})}\BibitemShut {NoStop}%
\bibitem [{\citenamefont {Unruh}(1981)}]{Unruh}%
  \BibitemOpen
  \bibfield  {author} {\bibinfo {author} {\bibfnamefont {W.~G.}\ \bibnamefont
  {Unruh}},\ }\bibfield  {title} {\enquote {\bibinfo {title} {{Experimental
  Black-Hole Evaporation?}}}\ }\href {\doibase 10.1103/PhysRevLett.46.1351}
  {\bibfield  {journal} {\bibinfo  {journal} {Phys. Rev. Lett.}\ }\textbf
  {\bibinfo {volume} {46}},\ \bibinfo {pages} {1351--1353} (\bibinfo {year}
  {1981})}\BibitemShut {NoStop}%
\bibitem [{\citenamefont {Fischer}\ and\ \citenamefont
  {Visser}(2003)}]{FischerVisser}%
  \BibitemOpen
  \bibfield  {author} {\bibinfo {author} {\bibfnamefont {Uwe~R.}\ \bibnamefont
  {Fischer}}\ and\ \bibinfo {author} {\bibfnamefont {Matt}\ \bibnamefont
  {Visser}},\ }\bibfield  {title} {\enquote {\bibinfo {title} {{On the
  space-time curvature experienced by quasiparticle excitations in the
  Painlev{\'e}--Gullstrand effective geometry}},}\ }\href {\doibase
  https://doi.org/10.1016/S0003-4916(03)00011-3} {\bibfield  {journal}
  {\bibinfo  {journal} {Annals of Physics}\ }\textbf {\bibinfo {volume}
  {304}},\ \bibinfo {pages} {22--39} (\bibinfo {year} {2003})}\BibitemShut
  {NoStop}%
\bibitem [{\citenamefont {Barcel\'o}\ \emph {et~al.}(2011)\citenamefont
  {Barcel\'o}, \citenamefont {Liberati},\ and\ \citenamefont {Visser}}]{BLV}%
  \BibitemOpen
  \bibfield  {author} {\bibinfo {author} {\bibfnamefont {C.}~\bibnamefont
  {Barcel\'o}}, \bibinfo {author} {\bibfnamefont {S.}~\bibnamefont {Liberati}},
  \ and\ \bibinfo {author} {\bibfnamefont {M.}~\bibnamefont {Visser}},\
  }\bibfield  {title} {\enquote {\bibinfo {title} {{Analogue Gravity}},}\
  }\href {\doibase 10.12942/lrr-2011-3} {\bibfield  {journal} {\bibinfo
  {journal} {Living Reviews in Relativity}\ }\textbf {\bibinfo {volume} {14}}
  (\bibinfo {year} {2011}),\ 10.12942/lrr-2011-3}\BibitemShut {NoStop}%
\bibitem [{\citenamefont {Fedichev}\ and\ \citenamefont {Fischer}(2004)}]{CPP}%
  \BibitemOpen
  \bibfield  {author} {\bibinfo {author} {\bibfnamefont {P.~O.}\ \bibnamefont
  {Fedichev}}\ and\ \bibinfo {author} {\bibfnamefont {U.~R.}\ \bibnamefont
  {Fischer}},\ }\bibfield  {title} {\enquote {\bibinfo {title}
  {{``Cosmological'' quasiparticle production in harmonically trapped
  superfluid gases}},}\ }\href {\doibase 10.1103/PhysRevA.69.033602} {\bibfield
   {journal} {\bibinfo  {journal} {Phys. Rev. A}\ }\textbf {\bibinfo {volume}
  {69}},\ \bibinfo {pages} {033602} (\bibinfo {year} {2004})}\BibitemShut
  {NoStop}%
\bibitem [{\citenamefont {Sch\"utzhold}\ \emph {et~al.}(2007)\citenamefont
  {Sch\"utzhold}, \citenamefont {Uhlmann}, \citenamefont {Petersen},
  \citenamefont {Schmitz}, \citenamefont {Friedenauer},\ and\ \citenamefont
  {Sch\"atz}}]{Schaetz}%
  \BibitemOpen
  \bibfield  {author} {\bibinfo {author} {\bibfnamefont {Ralf}\ \bibnamefont
  {Sch\"utzhold}}, \bibinfo {author} {\bibfnamefont {Michael}\ \bibnamefont
  {Uhlmann}}, \bibinfo {author} {\bibfnamefont {Lutz}\ \bibnamefont
  {Petersen}}, \bibinfo {author} {\bibfnamefont {Hector}\ \bibnamefont
  {Schmitz}}, \bibinfo {author} {\bibfnamefont {Axel}\ \bibnamefont
  {Friedenauer}}, \ and\ \bibinfo {author} {\bibfnamefont {Tobias}\
  \bibnamefont {Sch\"atz}},\ }\bibfield  {title} {\enquote {\bibinfo {title}
  {{Analogue of Cosmological Particle Creation in an Ion Trap}},}\ }\href
  {\doibase 10.1103/PhysRevLett.99.201301} {\bibfield  {journal} {\bibinfo
  {journal} {Phys. Rev. Lett.}\ }\textbf {\bibinfo {volume} {99}},\ \bibinfo
  {pages} {201301} (\bibinfo {year} {2007})}\BibitemShut {NoStop}%
\bibitem [{\citenamefont {Carusotto}\ \emph {et~al.}(2010)\citenamefont
  {Carusotto}, \citenamefont {Balbinot}, \citenamefont {Fabbri},\ and\
  \citenamefont {Recati}}]{Carusotto2010}%
  \BibitemOpen
  \bibfield  {author} {\bibinfo {author} {\bibfnamefont {I.}~\bibnamefont
  {Carusotto}}, \bibinfo {author} {\bibfnamefont {R.}~\bibnamefont {Balbinot}},
  \bibinfo {author} {\bibfnamefont {A.}~\bibnamefont {Fabbri}}, \ and\ \bibinfo
  {author} {\bibfnamefont {A.}~\bibnamefont {Recati}},\ }\bibfield  {title}
  {\enquote {\bibinfo {title} {{{Density correlations and analog dynamical
  Casimir emission of Bogoliubov phonons in modulated atomic Bose-Einstein
  condensates}}},}\ }\href {\doibase 10.1140/epjd/e2009-00314-3} {\bibfield
  {journal} {\bibinfo  {journal} {The European Physical Journal D}\ }\textbf
  {\bibinfo {volume} {56}},\ \bibinfo {pages} {391--404} (\bibinfo {year}
  {2010})}\BibitemShut {NoStop}%
\bibitem [{\citenamefont {Robertson}\ \emph
  {et~al.}(2017{\natexlab{a}})\citenamefont {Robertson}, \citenamefont
  {Michel},\ and\ \citenamefont {Parentani}}]{PhysRevD.95.065020}%
  \BibitemOpen
  \bibfield  {author} {\bibinfo {author} {\bibfnamefont {Scott}\ \bibnamefont
  {Robertson}}, \bibinfo {author} {\bibfnamefont {Florent}\ \bibnamefont
  {Michel}}, \ and\ \bibinfo {author} {\bibfnamefont {Renaud}\ \bibnamefont
  {Parentani}},\ }\bibfield  {title} {\enquote {\bibinfo {title} {{Controlling
  and observing nonseparability of phonons created in time-dependent 1D atomic
  Bose condensate}s},}\ }\href {\doibase 10.1103/PhysRevD.95.065020} {\bibfield
   {journal} {\bibinfo  {journal} {Phys. Rev. D}\ }\textbf {\bibinfo {volume}
  {95}},\ \bibinfo {pages} {065020} (\bibinfo {year}
  {2017}{\natexlab{a}})}\BibitemShut {NoStop}%
\bibitem [{\citenamefont {Johansson}\ \emph {et~al.}(2009)\citenamefont
  {Johansson}, \citenamefont {Johansson}, \citenamefont {Wilson},\ and\
  \citenamefont {Nori}}]{PhysRevLett.103.147003}%
  \BibitemOpen
  \bibfield  {author} {\bibinfo {author} {\bibfnamefont {J.~R.}\ \bibnamefont
  {Johansson}}, \bibinfo {author} {\bibfnamefont {G.}~\bibnamefont
  {Johansson}}, \bibinfo {author} {\bibfnamefont {C.~M.}\ \bibnamefont
  {Wilson}}, \ and\ \bibinfo {author} {\bibfnamefont {Franco}\ \bibnamefont
  {Nori}},\ }\bibfield  {title} {\enquote {\bibinfo {title} {{Dynamical Casimir
  Effect in a Superconducting Coplanar Waveguide}},}\ }\href {\doibase
  10.1103/PhysRevLett.103.147003} {\bibfield  {journal} {\bibinfo  {journal}
  {Phys. Rev. Lett.}\ }\textbf {\bibinfo {volume} {103}},\ \bibinfo {pages}
  {147003} (\bibinfo {year} {2009})}\BibitemShut {NoStop}%
\bibitem [{\citenamefont {Tian}\ \emph {et~al.}(2017)\citenamefont {Tian},
  \citenamefont {Jing},\ and\ \citenamefont {Dragan}}]{PhysRevD.95.125003}%
  \BibitemOpen
  \bibfield  {author} {\bibinfo {author} {\bibfnamefont {Zehua}\ \bibnamefont
  {Tian}}, \bibinfo {author} {\bibfnamefont {Jiliang}\ \bibnamefont {Jing}}, \
  and\ \bibinfo {author} {\bibfnamefont {Andrzej}\ \bibnamefont {Dragan}},\
  }\bibfield  {title} {\enquote {\bibinfo {title} {Analog cosmological particle
  generation in a superconducting circuit},}\ }\href {\doibase
  10.1103/PhysRevD.95.125003} {\bibfield  {journal} {\bibinfo  {journal} {Phys.
  Rev. D}\ }\textbf {\bibinfo {volume} {95}},\ \bibinfo {pages} {125003}
  (\bibinfo {year} {2017})}\BibitemShut {NoStop}%
\bibitem [{\citenamefont {Wilson}\ \emph {et~al.}(2011)\citenamefont {Wilson},
  \citenamefont {Johansson}, \citenamefont {Pourkabirian}, \citenamefont
  {Simoen}, \citenamefont {Johansson}, \citenamefont {Duty}, \citenamefont
  {Nori},\ and\ \citenamefont {Delsing}}]{DCE}%
  \BibitemOpen
  \bibfield  {author} {\bibinfo {author} {\bibfnamefont {C.~M.}\ \bibnamefont
  {Wilson}}, \bibinfo {author} {\bibfnamefont {G.}~\bibnamefont {Johansson}},
  \bibinfo {author} {\bibfnamefont {A.}~\bibnamefont {Pourkabirian}}, \bibinfo
  {author} {\bibfnamefont {M.}~\bibnamefont {Simoen}}, \bibinfo {author}
  {\bibfnamefont {J.~R.}\ \bibnamefont {Johansson}}, \bibinfo {author}
  {\bibfnamefont {T.}~\bibnamefont {Duty}}, \bibinfo {author} {\bibfnamefont
  {F.}~\bibnamefont {Nori}}, \ and\ \bibinfo {author} {\bibfnamefont
  {P.}~\bibnamefont {Delsing}},\ }\bibfield  {title} {\enquote {\bibinfo
  {title} {{Observation of the dynamical Casimir effect in a superconducting
  circuit}},}\ }\href {http://dx.doi.org/10.1038/nature10561} {\bibfield
  {journal} {\bibinfo  {journal} {Nature}\ }\textbf {\bibinfo {volume} {479}},\
  \bibinfo {pages} {376--379} (\bibinfo {year} {2011})}\BibitemShut {NoStop}%
\bibitem [{\citenamefont {L\"ahteenm\"aki}\ \emph {et~al.}(2013)\citenamefont
  {L\"ahteenm\"aki}, \citenamefont {Paraoanu}, \citenamefont {Hassel},\ and\
  \citenamefont {Hakonen}}]{Lahteenmaki}%
  \BibitemOpen
  \bibfield  {author} {\bibinfo {author} {\bibfnamefont {Pasi}\ \bibnamefont
  {L\"ahteenm\"aki}}, \bibinfo {author} {\bibfnamefont {G.~S.}\ \bibnamefont
  {Paraoanu}}, \bibinfo {author} {\bibfnamefont {Juha}\ \bibnamefont {Hassel}},
  \ and\ \bibinfo {author} {\bibfnamefont {Pertti~J.}\ \bibnamefont
  {Hakonen}},\ }\bibfield  {title} {\enquote {\bibinfo {title} {{Dynamical
  Casimir effect in a Josephson metamaterial}},}\ }\href {\doibase
  10.1073/pnas.1212705110} {\bibfield  {journal} {\bibinfo  {journal}
  {Proceedings of the National Academy of Sciences}\ }\textbf {\bibinfo
  {volume} {110}},\ \bibinfo {pages} {4234--4238} (\bibinfo {year}
  {2013})}\BibitemShut {NoStop}%
\bibitem [{\citenamefont {Jaskula}\ \emph {et~al.}(2012)\citenamefont
  {Jaskula}, \citenamefont {Partridge}, \citenamefont {Bonneau}, \citenamefont
  {Lopes}, \citenamefont {Ruaudel}, \citenamefont {Boiron},\ and\ \citenamefont
  {Westbrook}}]{PhysRevLett.109.220401}%
  \BibitemOpen
  \bibfield  {author} {\bibinfo {author} {\bibfnamefont {J.-C.}\ \bibnamefont
  {Jaskula}}, \bibinfo {author} {\bibfnamefont {G.~B.}\ \bibnamefont
  {Partridge}}, \bibinfo {author} {\bibfnamefont {M.}~\bibnamefont {Bonneau}},
  \bibinfo {author} {\bibfnamefont {R.}~\bibnamefont {Lopes}}, \bibinfo
  {author} {\bibfnamefont {J.}~\bibnamefont {Ruaudel}}, \bibinfo {author}
  {\bibfnamefont {D.}~\bibnamefont {Boiron}}, \ and\ \bibinfo {author}
  {\bibfnamefont {C.~I.}\ \bibnamefont {Westbrook}},\ }\bibfield  {title}
  {\enquote {\bibinfo {title} {{Acoustic Analog to the Dynamical Casimir Effect
  in a Bose-Einstein Condensate}},}\ }\href {\doibase
  10.1103/PhysRevLett.109.220401} {\bibfield  {journal} {\bibinfo  {journal}
  {Phys. Rev. Lett.}\ }\textbf {\bibinfo {volume} {109}},\ \bibinfo {pages}
  {220401} (\bibinfo {year} {2012})}\BibitemShut {NoStop}%
\bibitem [{\citenamefont {Hung}\ \emph {et~al.}(2013)\citenamefont {Hung},
  \citenamefont {Gurarie},\ and\ \citenamefont {Chin}}]{Chin}%
  \BibitemOpen
  \bibfield  {author} {\bibinfo {author} {\bibfnamefont {C.-L.}\ \bibnamefont
  {Hung}}, \bibinfo {author} {\bibfnamefont {V.}~\bibnamefont {Gurarie}}, \
  and\ \bibinfo {author} {\bibfnamefont {C.}~\bibnamefont {Chin}},\ }\bibfield
  {title} {\enquote {\bibinfo {title} {{From Cosmology to Cold Atoms:
  Observation of Sakharov Oscillations in a Quenched Atomic Superfluid}},}\
  }\href {\doibase 10.1126/science.1237557} {\bibfield  {journal} {\bibinfo
  {journal} {Science}\ }\textbf {\bibinfo {volume} {341}},\ \bibinfo {pages}
  {1213--1215} (\bibinfo {year} {2013})}\BibitemShut {NoStop}%
\bibitem [{\citenamefont {Visser}(1998{\natexlab{b}})}]{MattCQG}%
  \BibitemOpen
  \bibfield  {author} {\bibinfo {author} {\bibfnamefont {M.}~\bibnamefont
  {Visser}},\ }\bibfield  {title} {\enquote {\bibinfo {title} {{Acoustic black
  holes: horizons, ergospheres and Hawking radiation}},}\ }\href
  {http://stacks.iop.org/0264-9381/15/i=6/a=024} {\bibfield  {journal}
  {\bibinfo  {journal} {Classical and Quantum Gravity}\ }\textbf {\bibinfo
  {volume} {15}},\ \bibinfo {pages} {1767} (\bibinfo {year}
  {1998}{\natexlab{b}})}\BibitemShut {NoStop}%
\bibitem [{\citenamefont {Fischer}\ and\ \citenamefont
  {Sch\"utzhold}(2004)}]{Schuetzhold}%
  \BibitemOpen
  \bibfield  {author} {\bibinfo {author} {\bibfnamefont {U.~R.}\ \bibnamefont
  {Fischer}}\ and\ \bibinfo {author} {\bibfnamefont {R.}~\bibnamefont
  {Sch\"utzhold}},\ }\bibfield  {title} {\enquote {\bibinfo {title} {{Quantum
  simulation of cosmic inflation in two-component Bose-Einstein
  condensates}},}\ }\href {\doibase 10.1103/PhysRevA.70.063615} {\bibfield
  {journal} {\bibinfo  {journal} {Phys. Rev. A}\ }\textbf {\bibinfo {volume}
  {70}},\ \bibinfo {pages} {063615} (\bibinfo {year} {2004})}\BibitemShut
  {NoStop}%
\bibitem [{\citenamefont {Balbinot}\ \emph {et~al.}(2008)\citenamefont
  {Balbinot}, \citenamefont {Fabbri}, \citenamefont {Fagnocchi}, \citenamefont
  {Recati},\ and\ \citenamefont {Carusotto}}]{Balbinot}%
  \BibitemOpen
  \bibfield  {author} {\bibinfo {author} {\bibfnamefont {Roberto}\ \bibnamefont
  {Balbinot}}, \bibinfo {author} {\bibfnamefont {Alessandro}\ \bibnamefont
  {Fabbri}}, \bibinfo {author} {\bibfnamefont {Serena}\ \bibnamefont
  {Fagnocchi}}, \bibinfo {author} {\bibfnamefont {Alessio}\ \bibnamefont
  {Recati}}, \ and\ \bibinfo {author} {\bibfnamefont {Iacopo}\ \bibnamefont
  {Carusotto}},\ }\bibfield  {title} {\enquote {\bibinfo {title} {{Nonlocal
  density correlations as a signature of Hawking radiation from acoustic black
  holes}},}\ }\href {\doibase 10.1103/PhysRevA.78.021603} {\bibfield  {journal}
  {\bibinfo  {journal} {Phys. Rev. A}\ }\textbf {\bibinfo {volume} {78}},\
  \bibinfo {pages} {021603} (\bibinfo {year} {2008})}\BibitemShut {NoStop}%
\bibitem [{\citenamefont {Steinhauer}(2015)}]{PhysRevD.92.024043}%
  \BibitemOpen
  \bibfield  {author} {\bibinfo {author} {\bibfnamefont {Jeff}\ \bibnamefont
  {Steinhauer}},\ }\bibfield  {title} {\enquote {\bibinfo {title} {{Measuring
  the entanglement of analogue Hawking radiation by the density-density
  correlation function}},}\ }\href {\doibase 10.1103/PhysRevD.92.024043}
  {\bibfield  {journal} {\bibinfo  {journal} {Phys. Rev. D}\ }\textbf {\bibinfo
  {volume} {92}},\ \bibinfo {pages} {024043} (\bibinfo {year}
  {2015})}\BibitemShut {NoStop}%
\bibitem [{\citenamefont {Nation}\ \emph {et~al.}(2009)\citenamefont {Nation},
  \citenamefont {Blencowe}, \citenamefont {Rimberg},\ and\ \citenamefont
  {Buks}}]{PhysRevLett.103.087004}%
  \BibitemOpen
  \bibfield  {author} {\bibinfo {author} {\bibfnamefont {P.~D.}\ \bibnamefont
  {Nation}}, \bibinfo {author} {\bibfnamefont {M.~P.}\ \bibnamefont
  {Blencowe}}, \bibinfo {author} {\bibfnamefont {A.~J.}\ \bibnamefont
  {Rimberg}}, \ and\ \bibinfo {author} {\bibfnamefont {E.}~\bibnamefont
  {Buks}},\ }\bibfield  {title} {\enquote {\bibinfo {title} {{Analogue Hawking
  Radiation in a dc-SQUID Array Transmission Line}},}\ }\href {\doibase
  10.1103/PhysRevLett.103.087004} {\bibfield  {journal} {\bibinfo  {journal}
  {Phys. Rev. Lett.}\ }\textbf {\bibinfo {volume} {103}},\ \bibinfo {pages}
  {087004} (\bibinfo {year} {2009})}\BibitemShut {NoStop}%
\bibitem [{\citenamefont {de~Nova}\ \emph {et~al.}(2015)\citenamefont
  {de~Nova}, \citenamefont {Sols},\ and\ \citenamefont {Zapata}}]{NJP}%
  \BibitemOpen
  \bibfield  {author} {\bibinfo {author} {\bibfnamefont {J.~R.~M.}\
  \bibnamefont {de~Nova}}, \bibinfo {author} {\bibfnamefont {F.}~\bibnamefont
  {Sols}}, \ and\ \bibinfo {author} {\bibfnamefont {I.}~\bibnamefont
  {Zapata}},\ }\bibfield  {title} {\enquote {\bibinfo {title} {{Entanglement
  and violation of classical inequalities in the Hawking radiation of flowing
  atom condensates}},}\ }\href
  {http://stacks.iop.org/1367-2630/17/i=10/a=105003} {\bibfield  {journal}
  {\bibinfo  {journal} {New Journal of Physics}\ }\textbf {\bibinfo {volume}
  {17}},\ \bibinfo {pages} {105003} (\bibinfo {year} {2015})}\BibitemShut
  {NoStop}%
\bibitem [{\citenamefont {Robertson}\ \emph
  {et~al.}(2017{\natexlab{b}})\citenamefont {Robertson}, \citenamefont
  {Michel},\ and\ \citenamefont {Parentani}}]{PhysRevD.96.045012}%
  \BibitemOpen
  \bibfield  {author} {\bibinfo {author} {\bibfnamefont {Scott}\ \bibnamefont
  {Robertson}}, \bibinfo {author} {\bibfnamefont {Florent}\ \bibnamefont
  {Michel}}, \ and\ \bibinfo {author} {\bibfnamefont {Renaud}\ \bibnamefont
  {Parentani}},\ }\bibfield  {title} {\enquote {\bibinfo {title} {{Assessing
  degrees of entanglement of phonon states in atomic Bose gases through the
  measurement of commuting observables}},}\ }\href {\doibase
  10.1103/PhysRevD.96.045012} {\bibfield  {journal} {\bibinfo  {journal} {Phys.
  Rev. D}\ }\textbf {\bibinfo {volume} {96}},\ \bibinfo {pages} {045012}
  (\bibinfo {year} {2017}{\natexlab{b}})}\BibitemShut {NoStop}%
\bibitem [{\citenamefont {Nguyen}\ \emph {et~al.}(2015)\citenamefont {Nguyen},
  \citenamefont {Gerace}, \citenamefont {Carusotto}, \citenamefont {Sanvitto},
  \citenamefont {Galopin}, \citenamefont {Lema\^{\i}tre}, \citenamefont
  {Sagnes}, \citenamefont {Bloch},\ and\ \citenamefont
  {Amo}}]{PhysRevLett.114.036402}%
  \BibitemOpen
  \bibfield  {author} {\bibinfo {author} {\bibfnamefont {H.~S.}\ \bibnamefont
  {Nguyen}}, \bibinfo {author} {\bibfnamefont {D.}~\bibnamefont {Gerace}},
  \bibinfo {author} {\bibfnamefont {I.}~\bibnamefont {Carusotto}}, \bibinfo
  {author} {\bibfnamefont {D.}~\bibnamefont {Sanvitto}}, \bibinfo {author}
  {\bibfnamefont {E.}~\bibnamefont {Galopin}}, \bibinfo {author} {\bibfnamefont
  {A.}~\bibnamefont {Lema\^{\i}tre}}, \bibinfo {author} {\bibfnamefont
  {I.}~\bibnamefont {Sagnes}}, \bibinfo {author} {\bibfnamefont
  {J.}~\bibnamefont {Bloch}}, \ and\ \bibinfo {author} {\bibfnamefont
  {A.}~\bibnamefont {Amo}},\ }\bibfield  {title} {\enquote {\bibinfo {title}
  {{Acoustic Black Hole in a Stationary Hydrodynamic Flow of Microcavity
  Polaritons}},}\ }\href {\doibase 10.1103/PhysRevLett.114.036402} {\bibfield
  {journal} {\bibinfo  {journal} {Phys. Rev. Lett.}\ }\textbf {\bibinfo
  {volume} {114}},\ \bibinfo {pages} {036402} (\bibinfo {year}
  {2015})}\BibitemShut {NoStop}%
\bibitem [{\citenamefont {Lahav}\ \emph {et~al.}(2010)\citenamefont {Lahav},
  \citenamefont {Itah}, \citenamefont {Blumkin}, \citenamefont {Gordon},
  \citenamefont {Rinott}, \citenamefont {Zayats},\ and\ \citenamefont
  {Steinhauer}}]{PhysRevLett.105.240401}%
  \BibitemOpen
  \bibfield  {author} {\bibinfo {author} {\bibfnamefont {Oren}\ \bibnamefont
  {Lahav}}, \bibinfo {author} {\bibfnamefont {Amir}\ \bibnamefont {Itah}},
  \bibinfo {author} {\bibfnamefont {Alex}\ \bibnamefont {Blumkin}}, \bibinfo
  {author} {\bibfnamefont {Carmit}\ \bibnamefont {Gordon}}, \bibinfo {author}
  {\bibfnamefont {Shahar}\ \bibnamefont {Rinott}}, \bibinfo {author}
  {\bibfnamefont {Alona}\ \bibnamefont {Zayats}}, \ and\ \bibinfo {author}
  {\bibfnamefont {Jeff}\ \bibnamefont {Steinhauer}},\ }\bibfield  {title}
  {\enquote {\bibinfo {title} {{Realization of a Sonic Black Hole Analog in a
  Bose-Einstein Condensate}},}\ }\href {\doibase
  10.1103/PhysRevLett.105.240401} {\bibfield  {journal} {\bibinfo  {journal}
  {Phys. Rev. Lett.}\ }\textbf {\bibinfo {volume} {105}},\ \bibinfo {pages}
  {240401} (\bibinfo {year} {2010})}\BibitemShut {NoStop}%
\bibitem [{\citenamefont {Steinhauer}(2014)}]{BHlaser}%
  \BibitemOpen
  \bibfield  {author} {\bibinfo {author} {\bibfnamefont {Jeff}\ \bibnamefont
  {Steinhauer}},\ }\bibfield  {title} {\enquote {\bibinfo {title} {{Observation
  of self-amplifying Hawking radiation in an analogue black-hole laser}},}\
  }\href {http://dx.doi.org/10.1038/nphys3104} {\bibfield  {journal} {\bibinfo
  {journal} {Nat. Phys.}\ }\textbf {\bibinfo {volume} {10}},\ \bibinfo {pages}
  {864--869} (\bibinfo {year} {2014})}\BibitemShut {NoStop}%
\bibitem [{\citenamefont {Steinhauer}(2016)}]{QCHawking}%
  \BibitemOpen
  \bibfield  {author} {\bibinfo {author} {\bibfnamefont {Jeff}\ \bibnamefont
  {Steinhauer}},\ }\bibfield  {title} {\enquote {\bibinfo {title} {{Observation
  of quantum Hawking radiation and its entanglement in an analogue black
  hole}},}\ }\href {http://dx.doi.org/10.1038/nphys3863} {\bibfield  {journal}
  {\bibinfo  {journal} {Nat. Phys.}\ }\textbf {\bibinfo {volume} {12}},\
  \bibinfo {pages} {959--965} (\bibinfo {year} {2016})}\BibitemShut {NoStop}%
\bibitem [{\citenamefont {Eckel}\ \emph {et~al.}(2018)\citenamefont {Eckel},
  \citenamefont {Kumar}, \citenamefont {Jacobson}, \citenamefont {Spielman},\
  and\ \citenamefont {Campbell}}]{Gretchen}%
  \BibitemOpen
  \bibfield  {author} {\bibinfo {author} {\bibfnamefont {S.}~\bibnamefont
  {Eckel}}, \bibinfo {author} {\bibfnamefont {A.}~\bibnamefont {Kumar}},
  \bibinfo {author} {\bibfnamefont {T.}~\bibnamefont {Jacobson}}, \bibinfo
  {author} {\bibfnamefont {I.~B.}\ \bibnamefont {Spielman}}, \ and\ \bibinfo
  {author} {\bibfnamefont {G.~K.}\ \bibnamefont {Campbell}},\ }\bibfield
  {title} {\enquote {\bibinfo {title} {{A Rapidly Expanding Bose-Einstein
  Condensate: An Expanding Universe in the Lab}},}\ }\href {\doibase
  10.1103/PhysRevX.8.021021} {\bibfield  {journal} {\bibinfo  {journal} {Phys.
  Rev. X}\ }\textbf {\bibinfo {volume} {8}},\ \bibinfo {pages} {021021}
  (\bibinfo {year} {2018})}\BibitemShut {NoStop}%
\bibitem [{\citenamefont {Finazzi}\ and\ \citenamefont
  {Carusotto}(2014)}]{Finazzi}%
  \BibitemOpen
  \bibfield  {author} {\bibinfo {author} {\bibfnamefont {Stefano}\ \bibnamefont
  {Finazzi}}\ and\ \bibinfo {author} {\bibfnamefont {Iacopo}\ \bibnamefont
  {Carusotto}},\ }\bibfield  {title} {\enquote {\bibinfo {title} {{Entangled
  phonons in atomic Bose-Einstein condensates}},}\ }\href {\doibase
  10.1103/PhysRevA.90.033607} {\bibfield  {journal} {\bibinfo  {journal} {Phys.
  Rev. A}\ }\textbf {\bibinfo {volume} {90}},\ \bibinfo {pages} {033607}
  (\bibinfo {year} {2014})}\BibitemShut {NoStop}%
\bibitem [{\citenamefont {Baranov}(2008)}]{Baranov}%
  \BibitemOpen
  \bibfield  {author} {\bibinfo {author} {\bibfnamefont {M.~A.}\ \bibnamefont
  {Baranov}},\ }\bibfield  {title} {\enquote {\bibinfo {title} {Theoretical
  progress in many-body physics with ultracold dipolar gases},}\ }\href
  {\doibase https://doi.org/10.1016/j.physrep.2008.04.007} {\bibfield
  {journal} {\bibinfo  {journal} {Physics Reports}\ }\textbf {\bibinfo {volume}
  {464}},\ \bibinfo {pages} {71--111} (\bibinfo {year} {2008})}\BibitemShut
  {NoStop}%
\bibitem [{\citenamefont {Lahaye}\ \emph {et~al.}(2007)\citenamefont {Lahaye},
  \citenamefont {Koch}, \citenamefont {Fr\"ohlich}, \citenamefont {Fattori},
  \citenamefont {Metz}, \citenamefont {Griesmaier}, \citenamefont
  {Giovanazzi},\ and\ \citenamefont {Pfau}}]{Chromium}%
  \BibitemOpen
  \bibfield  {author} {\bibinfo {author} {\bibfnamefont {Thierry}\ \bibnamefont
  {Lahaye}}, \bibinfo {author} {\bibfnamefont {Tobias}\ \bibnamefont {Koch}},
  \bibinfo {author} {\bibfnamefont {Bernd}\ \bibnamefont {Fr\"ohlich}},
  \bibinfo {author} {\bibfnamefont {Marco}\ \bibnamefont {Fattori}}, \bibinfo
  {author} {\bibfnamefont {Jonas}\ \bibnamefont {Metz}}, \bibinfo {author}
  {\bibfnamefont {Axel}\ \bibnamefont {Griesmaier}}, \bibinfo {author}
  {\bibfnamefont {Stefano}\ \bibnamefont {Giovanazzi}}, \ and\ \bibinfo
  {author} {\bibfnamefont {Tilman}\ \bibnamefont {Pfau}},\ }\bibfield  {title}
  {\enquote {\bibinfo {title} {Strong dipolar effects in a quantum
  ferrofluid},}\ }\href {http://dx.doi.org/10.1038/nature06036} {\bibfield
  {journal} {\bibinfo  {journal} {Nature}\ }\textbf {\bibinfo {volume} {448}},\
  \bibinfo {pages} {672--675} (\bibinfo {year} {2007})}\BibitemShut {NoStop}%
\bibitem [{\citenamefont {Lu}\ \emph {et~al.}(2011)\citenamefont {Lu},
  \citenamefont {Burdick}, \citenamefont {Youn},\ and\ \citenamefont
  {Lev}}]{PhysRevLett.107.190401}%
  \BibitemOpen
  \bibfield  {author} {\bibinfo {author} {\bibfnamefont {Mingwu}\ \bibnamefont
  {Lu}}, \bibinfo {author} {\bibfnamefont {Nathaniel~Q.}\ \bibnamefont
  {Burdick}}, \bibinfo {author} {\bibfnamefont {Seo~Ho}\ \bibnamefont {Youn}},
  \ and\ \bibinfo {author} {\bibfnamefont {Benjamin~L.}\ \bibnamefont {Lev}},\
  }\bibfield  {title} {\enquote {\bibinfo {title} {{Strongly Dipolar
  Bose-Einstein Condensate of Dysprosium}},}\ }\href {\doibase
  10.1103/PhysRevLett.107.190401} {\bibfield  {journal} {\bibinfo  {journal}
  {Phys. Rev. Lett.}\ }\textbf {\bibinfo {volume} {107}},\ \bibinfo {pages}
  {190401} (\bibinfo {year} {2011})}\BibitemShut {NoStop}%
\bibitem [{\citenamefont {Aikawa}\ \emph {et~al.}(2012)\citenamefont {Aikawa},
  \citenamefont {Frisch}, \citenamefont {Mark}, \citenamefont {Baier},
  \citenamefont {Rietzler}, \citenamefont {Grimm},\ and\ \citenamefont
  {Ferlaino}}]{PhysRevLett.108.210401}%
  \BibitemOpen
  \bibfield  {author} {\bibinfo {author} {\bibfnamefont {K.}~\bibnamefont
  {Aikawa}}, \bibinfo {author} {\bibfnamefont {A.}~\bibnamefont {Frisch}},
  \bibinfo {author} {\bibfnamefont {M.}~\bibnamefont {Mark}}, \bibinfo {author}
  {\bibfnamefont {S.}~\bibnamefont {Baier}}, \bibinfo {author} {\bibfnamefont
  {A.}~\bibnamefont {Rietzler}}, \bibinfo {author} {\bibfnamefont
  {R.}~\bibnamefont {Grimm}}, \ and\ \bibinfo {author} {\bibfnamefont
  {F.}~\bibnamefont {Ferlaino}},\ }\bibfield  {title} {\enquote {\bibinfo
  {title} {{Bose-Einstein Condensation of Erbium}},}\ }\href {\doibase
  10.1103/PhysRevLett.108.210401} {\bibfield  {journal} {\bibinfo  {journal}
  {Phys. Rev. Lett.}\ }\textbf {\bibinfo {volume} {108}},\ \bibinfo {pages}
  {210401} (\bibinfo {year} {2012})}\BibitemShut {NoStop}%
\bibitem [{\citenamefont {Qu\'em\'ener}\ and\ \citenamefont
  {Julienne}(2012)}]{doi:10.1021/cr300092g}%
  \BibitemOpen
  \bibfield  {author} {\bibinfo {author} {\bibfnamefont {Goulven}\ \bibnamefont
  {Qu\'em\'ener}}\ and\ \bibinfo {author} {\bibfnamefont {Paul~S.}\
  \bibnamefont {Julienne}},\ }\bibfield  {title} {\enquote {\bibinfo {title}
  {{Ultracold Molecules under Control!}}}\ }\href {\doibase 10.1021/cr300092g}
  {\bibfield  {journal} {\bibinfo  {journal} {Chemical Reviews}\ }\textbf
  {\bibinfo {volume} {112}},\ \bibinfo {pages} {4949--5011} (\bibinfo {year}
  {2012})}\BibitemShut {NoStop}%
\bibitem [{\citenamefont {Guo}\ \emph {et~al.}(2016)\citenamefont {Guo},
  \citenamefont {Zhu}, \citenamefont {Lu}, \citenamefont {Ye}, \citenamefont
  {Wang}, \citenamefont {Vexiau}, \citenamefont {Bouloufa-Maafa}, \citenamefont
  {Qu\'em\'ener}, \citenamefont {Dulieu},\ and\ \citenamefont {Wang}}]{Guo}%
  \BibitemOpen
  \bibfield  {author} {\bibinfo {author} {\bibfnamefont {Mingyang}\
  \bibnamefont {Guo}}, \bibinfo {author} {\bibfnamefont {Bing}\ \bibnamefont
  {Zhu}}, \bibinfo {author} {\bibfnamefont {Bo}~\bibnamefont {Lu}}, \bibinfo
  {author} {\bibfnamefont {Xin}\ \bibnamefont {Ye}}, \bibinfo {author}
  {\bibfnamefont {Fudong}\ \bibnamefont {Wang}}, \bibinfo {author}
  {\bibfnamefont {Romain}\ \bibnamefont {Vexiau}}, \bibinfo {author}
  {\bibfnamefont {Nadia}\ \bibnamefont {Bouloufa-Maafa}}, \bibinfo {author}
  {\bibfnamefont {Goulven}\ \bibnamefont {Qu\'em\'ener}}, \bibinfo {author}
  {\bibfnamefont {Olivier}\ \bibnamefont {Dulieu}}, \ and\ \bibinfo {author}
  {\bibfnamefont {Dajun}\ \bibnamefont {Wang}},\ }\bibfield  {title} {\enquote
  {\bibinfo {title} {Creation of an ultracold gas of ground-state dipolar
  $^{23}\mathrm{Na}^{87}\mathrm{Rb}$ molecules},}\ }\href {\doibase
  10.1103/PhysRevLett.116.205303} {\bibfield  {journal} {\bibinfo  {journal}
  {Phys. Rev. Lett.}\ }\textbf {\bibinfo {volume} {116}},\ \bibinfo {pages}
  {205303} (\bibinfo {year} {2016})}\BibitemShut {NoStop}%
\bibitem [{\citenamefont {Rvachov}\ \emph {et~al.}(2017)\citenamefont
  {Rvachov}, \citenamefont {Son}, \citenamefont {Sommer}, \citenamefont
  {Ebadi}, \citenamefont {Park}, \citenamefont {Zwierlein}, \citenamefont
  {Ketterle},\ and\ \citenamefont {Jamison}}]{Rvachov}%
  \BibitemOpen
  \bibfield  {author} {\bibinfo {author} {\bibfnamefont {Timur~M.}\
  \bibnamefont {Rvachov}}, \bibinfo {author} {\bibfnamefont {Hyungmok}\
  \bibnamefont {Son}}, \bibinfo {author} {\bibfnamefont {Ariel~T.}\
  \bibnamefont {Sommer}}, \bibinfo {author} {\bibfnamefont {Sepehr}\
  \bibnamefont {Ebadi}}, \bibinfo {author} {\bibfnamefont {Juliana~J.}\
  \bibnamefont {Park}}, \bibinfo {author} {\bibfnamefont {Martin~W.}\
  \bibnamefont {Zwierlein}}, \bibinfo {author} {\bibfnamefont {Wolfgang}\
  \bibnamefont {Ketterle}}, \ and\ \bibinfo {author} {\bibfnamefont {Alan~O.}\
  \bibnamefont {Jamison}},\ }\bibfield  {title} {\enquote {\bibinfo {title}
  {Long-lived ultracold molecules with electric and magnetic dipole moments},}\
  }\href {\doibase 10.1103/PhysRevLett.119.143001} {\bibfield  {journal}
  {\bibinfo  {journal} {Phys. Rev. Lett.}\ }\textbf {\bibinfo {volume} {119}},\
  \bibinfo {pages} {143001} (\bibinfo {year} {2017})}\BibitemShut {NoStop}%
\bibitem [{\citenamefont {Santos}\ \emph {et~al.}(2003)\citenamefont {Santos},
  \citenamefont {Shlyapnikov},\ and\ \citenamefont
  {Lewenstein}}]{PhysRevLett.90.250403}%
  \BibitemOpen
  \bibfield  {author} {\bibinfo {author} {\bibfnamefont {L.}~\bibnamefont
  {Santos}}, \bibinfo {author} {\bibfnamefont {G.~V.}\ \bibnamefont
  {Shlyapnikov}}, \ and\ \bibinfo {author} {\bibfnamefont {M.}~\bibnamefont
  {Lewenstein}},\ }\bibfield  {title} {\enquote {\bibinfo {title} {{Roton-Maxon
  Spectrum and Stability of Trapped Dipolar Bose-Einstein Condensates}},}\
  }\href {\doibase 10.1103/PhysRevLett.90.250403} {\bibfield  {journal}
  {\bibinfo  {journal} {Phys. Rev. Lett.}\ }\textbf {\bibinfo {volume} {90}},\
  \bibinfo {pages} {250403} (\bibinfo {year} {2003})}\BibitemShut {NoStop}%
\bibitem [{\citenamefont {Fischer}(2006)}]{PhysRevA.73.031602}%
  \BibitemOpen
  \bibfield  {author} {\bibinfo {author} {\bibfnamefont {Uwe~R.}\ \bibnamefont
  {Fischer}},\ }\bibfield  {title} {\enquote {\bibinfo {title} {{Stability of
  quasi-two-dimensional Bose-Einstein condensates with dominant dipole-dipole
  interactions}},}\ }\href {\doibase 10.1103/PhysRevA.73.031602} {\bibfield
  {journal} {\bibinfo  {journal} {Phys. Rev. A}\ }\textbf {\bibinfo {volume}
  {73}},\ \bibinfo {pages} {031602} (\bibinfo {year} {2006})}\BibitemShut
  {NoStop}%
\bibitem [{\citenamefont {Ronen}\ \emph {et~al.}(2007)\citenamefont {Ronen},
  \citenamefont {Bortolotti},\ and\ \citenamefont
  {Bohn}}]{PhysRevLett.98.030406}%
  \BibitemOpen
  \bibfield  {author} {\bibinfo {author} {\bibfnamefont {Shai}\ \bibnamefont
  {Ronen}}, \bibinfo {author} {\bibfnamefont {Daniele C.~E.}\ \bibnamefont
  {Bortolotti}}, \ and\ \bibinfo {author} {\bibfnamefont {John~L.}\
  \bibnamefont {Bohn}},\ }\bibfield  {title} {\enquote {\bibinfo {title}
  {{Radial and Angular Rotons in Trapped Dipolar Gases}},}\ }\href {\doibase
  10.1103/PhysRevLett.98.030406} {\bibfield  {journal} {\bibinfo  {journal}
  {Phys. Rev. Lett.}\ }\textbf {\bibinfo {volume} {98}},\ \bibinfo {pages}
  {030406} (\bibinfo {year} {2007})}\BibitemShut {NoStop}%
\bibitem [{\citenamefont {Natu}\ \emph {et~al.}(2014)\citenamefont {Natu},
  \citenamefont {Campanello},\ and\ \citenamefont
  {Das~Sarma}}]{PhysRevA.90.043617}%
  \BibitemOpen
  \bibfield  {author} {\bibinfo {author} {\bibfnamefont {Stefan~S.}\
  \bibnamefont {Natu}}, \bibinfo {author} {\bibfnamefont {L.}~\bibnamefont
  {Campanello}}, \ and\ \bibinfo {author} {\bibfnamefont {S.}~\bibnamefont
  {Das~Sarma}},\ }\bibfield  {title} {\enquote {\bibinfo {title} {{Dynamics of
  correlations in a quasi-two-dimensional dipolar Bose gas following a quantum
  quench}},}\ }\href {\doibase 10.1103/PhysRevA.90.043617} {\bibfield
  {journal} {\bibinfo  {journal} {Phys. Rev. A}\ }\textbf {\bibinfo {volume}
  {90}},\ \bibinfo {pages} {043617} (\bibinfo {year} {2014})}\BibitemShut
  {NoStop}%
\bibitem [{\citenamefont {Ch\"a}\ and\ \citenamefont
  {Fischer}(2017)}]{PhysRevLett.118.130404}%
  \BibitemOpen
  \bibfield  {author} {\bibinfo {author} {\bibfnamefont {Seok-Yeong}\
  \bibnamefont {Ch\"a}}\ and\ \bibinfo {author} {\bibfnamefont {Uwe~R.}\
  \bibnamefont {Fischer}},\ }\bibfield  {title} {\enquote {\bibinfo {title}
  {{Probing the Scale Invariance of the Inflationary Power Spectrum in
  Expanding Quasi-Two-Dimensional Dipolar Condensates}},}\ }\href {\doibase
  10.1103/PhysRevLett.118.130404} {\bibfield  {journal} {\bibinfo  {journal}
  {Phys. Rev. Lett.}\ }\textbf {\bibinfo {volume} {118}},\ \bibinfo {pages}
  {130404} (\bibinfo {year} {2017})}\BibitemShut {NoStop}%
\bibitem [{\citenamefont {Landau}(1949)}]{Landau}%
  \BibitemOpen
  \bibfield  {author} {\bibinfo {author} {\bibfnamefont {L.}~\bibnamefont
  {Landau}},\ }\bibfield  {title} {\enquote {\bibinfo {title} {{On the Theory
  of Superfluidity}},}\ }\href {\doibase 10.1103/PhysRev.75.884} {\bibfield
  {journal} {\bibinfo  {journal} {Phys. Rev.}\ }\textbf {\bibinfo {volume}
  {75}},\ \bibinfo {pages} {884--885} (\bibinfo {year} {1949})}\BibitemShut
  {NoStop}%
\bibitem [{\citenamefont {Mezei}(1980)}]{Mezei}%
  \BibitemOpen
  \bibfield  {author} {\bibinfo {author} {\bibfnamefont {F.}~\bibnamefont
  {Mezei}},\ }\bibfield  {title} {\enquote {\bibinfo {title} {{High-Resolution
  Study of Excitations in Superfluid $^{4}\mathrm{He}$ by the Neutron Spin-Echo
  Technique}},}\ }\href {\doibase 10.1103/PhysRevLett.44.1601} {\bibfield
  {journal} {\bibinfo  {journal} {Phys. Rev. Lett.}\ }\textbf {\bibinfo
  {volume} {44}},\ \bibinfo {pages} {1601--1604} (\bibinfo {year}
  {1980})}\BibitemShut {NoStop}%
\bibitem [{\citenamefont {Kadau}\ \emph {et~al.}(2016)\citenamefont {Kadau},
  \citenamefont {Schmitt}, \citenamefont {Wenzel}, \citenamefont {Wink},
  \citenamefont {Maier}, \citenamefont {Ferrier-Barbut},\ and\ \citenamefont
  {Pfau}}]{DDIexperiment}%
  \BibitemOpen
  \bibfield  {author} {\bibinfo {author} {\bibfnamefont {Holger}\ \bibnamefont
  {Kadau}}, \bibinfo {author} {\bibfnamefont {Matthias}\ \bibnamefont
  {Schmitt}}, \bibinfo {author} {\bibfnamefont {Matthias}\ \bibnamefont
  {Wenzel}}, \bibinfo {author} {\bibfnamefont {Clarissa}\ \bibnamefont {Wink}},
  \bibinfo {author} {\bibfnamefont {Thomas}\ \bibnamefont {Maier}}, \bibinfo
  {author} {\bibfnamefont {Igor}\ \bibnamefont {Ferrier-Barbut}}, \ and\
  \bibinfo {author} {\bibfnamefont {Tilman}\ \bibnamefont {Pfau}},\ }\bibfield
  {title} {\enquote {\bibinfo {title} {{Observing the Rosensweig instability of
  a quantum ferrofluid}},}\ }\href {http://dx.doi.org/10.1038/nature16485}
  {\bibfield  {journal} {\bibinfo  {journal} {Nature}\ }\textbf {\bibinfo
  {volume} {530}},\ \bibinfo {pages} {194--197} (\bibinfo {year}
  {2016})}\BibitemShut {NoStop}%
\bibitem [{\citenamefont {Ferrier-Barbut}\ \emph {et~al.}(2016)\citenamefont
  {Ferrier-Barbut}, \citenamefont {Kadau}, \citenamefont {Schmitt},
  \citenamefont {Wenzel},\ and\ \citenamefont {Pfau}}]{PhysRevLett.116.215301}%
  \BibitemOpen
  \bibfield  {author} {\bibinfo {author} {\bibfnamefont {Igor}\ \bibnamefont
  {Ferrier-Barbut}}, \bibinfo {author} {\bibfnamefont {Holger}\ \bibnamefont
  {Kadau}}, \bibinfo {author} {\bibfnamefont {Matthias}\ \bibnamefont
  {Schmitt}}, \bibinfo {author} {\bibfnamefont {Matthias}\ \bibnamefont
  {Wenzel}}, \ and\ \bibinfo {author} {\bibfnamefont {Tilman}\ \bibnamefont
  {Pfau}},\ }\bibfield  {title} {\enquote {\bibinfo {title} {{Observation of
  Quantum Droplets in a Strongly Dipolar Bose Gas}},}\ }\href {\doibase
  10.1103/PhysRevLett.116.215301} {\bibfield  {journal} {\bibinfo  {journal}
  {Phys. Rev. Lett.}\ }\textbf {\bibinfo {volume} {116}},\ \bibinfo {pages}
  {215301} (\bibinfo {year} {2016})}\BibitemShut {NoStop}%
\bibitem [{\citenamefont {Chomaz}\ \emph {et~al.}(2018)\citenamefont {Chomaz},
  \citenamefont {van Bijnen}, \citenamefont {Petter}, \citenamefont {Faraoni},
  \citenamefont {Baier}, \citenamefont {Becher}, \citenamefont {Mark},
  \citenamefont {W{\"a}chtler}, \citenamefont {Santos},\ and\ \citenamefont
  {Ferlaino}}]{Chomaz}%
  \BibitemOpen
  \bibfield  {author} {\bibinfo {author} {\bibfnamefont {L.}~\bibnamefont
  {Chomaz}}, \bibinfo {author} {\bibfnamefont {R.~M.~W.}\ \bibnamefont {van
  Bijnen}}, \bibinfo {author} {\bibfnamefont {D.}~\bibnamefont {Petter}},
  \bibinfo {author} {\bibfnamefont {G.}~\bibnamefont {Faraoni}}, \bibinfo
  {author} {\bibfnamefont {S.}~\bibnamefont {Baier}}, \bibinfo {author}
  {\bibfnamefont {J.~H.}\ \bibnamefont {Becher}}, \bibinfo {author}
  {\bibfnamefont {M.~J.}\ \bibnamefont {Mark}}, \bibinfo {author}
  {\bibfnamefont {F.}~\bibnamefont {W{\"a}chtler}}, \bibinfo {author}
  {\bibfnamefont {L.}~\bibnamefont {Santos}}, \ and\ \bibinfo {author}
  {\bibfnamefont {F.}~\bibnamefont {Ferlaino}},\ }\bibfield  {title} {\enquote
  {\bibinfo {title} {Observation of roton mode population in a dipolar quantum
  gas},}\ }\href {\doibase 10.1038/s41567-018-0054-7} {\bibfield  {journal}
  {\bibinfo  {journal} {Nature Physics}\ }\textbf {\bibinfo {volume} {14}},\
  \bibinfo {pages} {442--446} (\bibinfo {year} {2018})}\BibitemShut {NoStop}%
\bibitem [{\citenamefont {Wenzel}\ \emph {et~al.}(2017)\citenamefont {Wenzel},
  \citenamefont {B\"ottcher}, \citenamefont {Langen}, \citenamefont
  {Ferrier-Barbut},\ and\ \citenamefont {Pfau}}]{Wenzel}%
  \BibitemOpen
  \bibfield  {author} {\bibinfo {author} {\bibfnamefont {Matthias}\
  \bibnamefont {Wenzel}}, \bibinfo {author} {\bibfnamefont {Fabian}\
  \bibnamefont {B\"ottcher}}, \bibinfo {author} {\bibfnamefont {Tim}\
  \bibnamefont {Langen}}, \bibinfo {author} {\bibfnamefont {Igor}\ \bibnamefont
  {Ferrier-Barbut}}, \ and\ \bibinfo {author} {\bibfnamefont {Tilman}\
  \bibnamefont {Pfau}},\ }\bibfield  {title} {\enquote {\bibinfo {title}
  {Striped states in a many-body system of tilted dipoles},}\ }\href {\doibase
  10.1103/PhysRevA.96.053630} {\bibfield  {journal} {\bibinfo  {journal} {Phys.
  Rev. A}\ }\textbf {\bibinfo {volume} {96}},\ \bibinfo {pages} {053630}
  (\bibinfo {year} {2017})}\BibitemShut {NoStop}%
\bibitem [{\citenamefont {Kagan}\ \emph {et~al.}(1996)\citenamefont {Kagan},
  \citenamefont {Surkov},\ and\ \citenamefont
  {Shlyapnikov}}]{PhysRevA.54.R1753}%
  \BibitemOpen
  \bibfield  {author} {\bibinfo {author} {\bibfnamefont {Yu.}\ \bibnamefont
  {Kagan}}, \bibinfo {author} {\bibfnamefont {E.~L.}\ \bibnamefont {Surkov}}, \
  and\ \bibinfo {author} {\bibfnamefont {G.~V.}\ \bibnamefont {Shlyapnikov}},\
  }\bibfield  {title} {\enquote {\bibinfo {title} {{Evolution of a
  Bose-condensed gas under variations of the confining potential}},}\ }\href
  {\doibase 10.1103/PhysRevA.54.R1753} {\bibfield  {journal} {\bibinfo
  {journal} {Phys. Rev. A}\ }\textbf {\bibinfo {volume} {54}},\ \bibinfo
  {pages} {R1753--R1756} (\bibinfo {year} {1996})}\BibitemShut {NoStop}%
\bibitem [{\citenamefont {Castin}\ and\ \citenamefont
  {Dum}(1996)}]{PhysRevLett.77.5315}%
  \BibitemOpen
  \bibfield  {author} {\bibinfo {author} {\bibfnamefont {Y.}~\bibnamefont
  {Castin}}\ and\ \bibinfo {author} {\bibfnamefont {R.}~\bibnamefont {Dum}},\
  }\bibfield  {title} {\enquote {\bibinfo {title} {{Bose-Einstein Condensates
  in Time Dependent Traps}},}\ }\href {\doibase 10.1103/PhysRevLett.77.5315}
  {\bibfield  {journal} {\bibinfo  {journal} {Phys. Rev. Lett.}\ }\textbf
  {\bibinfo {volume} {77}},\ \bibinfo {pages} {5315--5319} (\bibinfo {year}
  {1996})}\BibitemShut {NoStop}%
\bibitem [{\citenamefont {Gritsev}\ \emph {et~al.}(2010)\citenamefont
  {Gritsev}, \citenamefont {Barmettler},\ and\ \citenamefont
  {Demler}}]{1367-2630-12-11-113005}%
  \BibitemOpen
  \bibfield  {author} {\bibinfo {author} {\bibfnamefont {Vladimir}\
  \bibnamefont {Gritsev}}, \bibinfo {author} {\bibfnamefont {Peter}\
  \bibnamefont {Barmettler}}, \ and\ \bibinfo {author} {\bibfnamefont {Eugene}\
  \bibnamefont {Demler}},\ }\bibfield  {title} {\enquote {\bibinfo {title}
  {Scaling approach to quantum non-equilibrium dynamics of many-body
  systems},}\ }\href {http://stacks.iop.org/1367-2630/12/i=11/a=113005}
  {\bibfield  {journal} {\bibinfo  {journal} {New Journal of Physics}\ }\textbf
  {\bibinfo {volume} {12}},\ \bibinfo {pages} {113005} (\bibinfo {year}
  {2010})}\BibitemShut {NoStop}%
\bibitem [{\citenamefont {{Castin}}(2001)}]{2001camw.book....1C}%
  \BibitemOpen
  \bibfield  {author} {\bibinfo {author} {\bibfnamefont {Y.}~\bibnamefont
  {{Castin}}},\ }\enquote {\bibinfo {title} {{{Bose-Einstein Condensates in
  Atomic Gases: Simple Theoretical Results}}},}\ in\ \href {\doibase
  10.1007/3-540-45338-5_1} {\emph {\bibinfo {booktitle} {Coherent atomic matter
  waves}}},\ \bibinfo {series and number} {Les Houches Session LXXII},\
  \bibinfo {editor} {edited by\ \bibinfo {editor} {\bibfnamefont
  {R.}~\bibnamefont {{Kaiser}}}, \bibinfo {editor} {\bibfnamefont
  {C.}~\bibnamefont {{Westbrook}}}, \ and\ \bibinfo {editor} {\bibfnamefont
  {F.}~\bibnamefont {{David}}}}\ (\bibinfo  {publisher} {Springer, Berlin},\
  \bibinfo {year} {2001})\ pp.\ \bibinfo {pages} {1--136}\BibitemShut {NoStop}%
\bibitem [{\citenamefont {Busch}\ \emph {et~al.}(2014)\citenamefont {Busch},
  \citenamefont {Parentani},\ and\ \citenamefont
  {Robertson}}]{PhysRevA.89.063606}%
  \BibitemOpen
  \bibfield  {author} {\bibinfo {author} {\bibfnamefont {Xavier}\ \bibnamefont
  {Busch}}, \bibinfo {author} {\bibfnamefont {Renaud}\ \bibnamefont
  {Parentani}}, \ and\ \bibinfo {author} {\bibfnamefont {Scott}\ \bibnamefont
  {Robertson}},\ }\bibfield  {title} {\enquote {\bibinfo {title} {{Quantum
  entanglement due to a modulated dynamical Casimir effect}},}\ }\href
  {\doibase 10.1103/PhysRevA.89.063606} {\bibfield  {journal} {\bibinfo
  {journal} {Phys. Rev. A}\ }\textbf {\bibinfo {volume} {89}},\ \bibinfo
  {pages} {063606} (\bibinfo {year} {2014})}\BibitemShut {NoStop}%
\bibitem [{\citenamefont {Hung}\ \emph {et~al.}(2011)\citenamefont {Hung},
  \citenamefont {Zhang}, \citenamefont {Ha}, \citenamefont {Tung},
  \citenamefont {Gemelke},\ and\ \citenamefont {Chin}}]{Hung}%
  \BibitemOpen
  \bibfield  {author} {\bibinfo {author} {\bibfnamefont {Chen-Lung}\
  \bibnamefont {Hung}}, \bibinfo {author} {\bibfnamefont {Xibo}\ \bibnamefont
  {Zhang}}, \bibinfo {author} {\bibfnamefont {Li-Chung}\ \bibnamefont {Ha}},
  \bibinfo {author} {\bibfnamefont {Shih-Kuang}\ \bibnamefont {Tung}}, \bibinfo
  {author} {\bibfnamefont {Nathan}\ \bibnamefont {Gemelke}}, \ and\ \bibinfo
  {author} {\bibfnamefont {Cheng}\ \bibnamefont {Chin}},\ }\bibfield  {title}
  {\enquote {\bibinfo {title} {Extracting density--density correlations from in
  situ images of atomic quantum gases},}\ }\href
  {http://stacks.iop.org/1367-2630/13/i=7/a=075019} {\bibfield  {journal}
  {\bibinfo  {journal} {New Journal of Physics}\ }\textbf {\bibinfo {volume}
  {13}},\ \bibinfo {pages} {075019} (\bibinfo {year} {2011})}\BibitemShut
  {NoStop}%
\bibitem [{Not()}]{NoteAsymptote}%
  \BibitemOpen
  \href@noop {} {\ }\bibinfo {note} {\!\!The function $\zeta w[\zeta]$
  occurring in $G_{2, \bk}$ approaches a constant in this limit
  \cite{PhysRevA.73.031602}.}\BibitemShut {Stop}%
\bibitem [{\citenamefont {Schr{\"o}dinger}(1935)}]{Schrödinger1935}%
  \BibitemOpen
  \bibfield  {author} {\bibinfo {author} {\bibfnamefont {E.}~\bibnamefont
  {Schr{\"o}dinger}},\ }\bibfield  {title} {\enquote {\bibinfo {title} {{Die
  gegenw{\"a}rtige Situation in der Quantenmechanik}},}\ }\href {\doibase
  10.1007/BF01491914} {\bibfield  {journal} {\bibinfo  {journal}
  {Naturwissenschaften}\ }\textbf {\bibinfo {volume} {23}},\ \bibinfo {pages}
  {823--828} (\bibinfo {year} {1935})}\BibitemShut {NoStop}%
\bibitem [{\citenamefont {Schr\"odinger}(1935)}]{schršoedinger_1935}%
  \BibitemOpen
  \bibfield  {author} {\bibinfo {author} {\bibfnamefont {E.}~\bibnamefont
  {Schr\"odinger}},\ }\bibfield  {title} {\enquote {\bibinfo {title}
  {{Discussion of Probability Relations between Separated Systems}},}\ }\href
  {\doibase 10.1017/S0305004100013554} {\bibfield  {journal} {\bibinfo
  {journal} {Mathematical Proceedings of the Cambridge Philosophical Society}\
  }\textbf {\bibinfo {volume} {31}},\ \bibinfo {pages} {555--563} (\bibinfo
  {year} {1935})}\BibitemShut {NoStop}%
\bibitem [{\citenamefont {Cavalcanti}\ and\ \citenamefont
  {Skrzypczyk}(2017)}]{Cavalcanti}%
  \BibitemOpen
  \bibfield  {author} {\bibinfo {author} {\bibfnamefont {D}~\bibnamefont
  {Cavalcanti}}\ and\ \bibinfo {author} {\bibfnamefont {P}~\bibnamefont
  {Skrzypczyk}},\ }\bibfield  {title} {\enquote {\bibinfo {title} {Quantum
  steering: a review with focus on semidefinite programming},}\ }\href
  {http://stacks.iop.org/0034-4885/80/i=2/a=024001} {\bibfield  {journal}
  {\bibinfo  {journal} {Reports on Progress in Physics}\ }\textbf {\bibinfo
  {volume} {80}},\ \bibinfo {pages} {024001} (\bibinfo {year}
  {2017})}\BibitemShut {NoStop}%
\bibitem [{\citenamefont {Wiseman}\ \emph {et~al.}(2007)\citenamefont
  {Wiseman}, \citenamefont {Jones},\ and\ \citenamefont
  {Doherty}}]{WisemanPRL}%
  \BibitemOpen
  \bibfield  {author} {\bibinfo {author} {\bibfnamefont {H.~M.}\ \bibnamefont
  {Wiseman}}, \bibinfo {author} {\bibfnamefont {S.~J.}\ \bibnamefont {Jones}},
  \ and\ \bibinfo {author} {\bibfnamefont {A.~C.}\ \bibnamefont {Doherty}},\
  }\bibfield  {title} {\enquote {\bibinfo {title} {{Steering, Entanglement,
  Nonlocality, and the Einstein-Podolsky-Rosen Paradox}},}\ }\href {\doibase
  10.1103/PhysRevLett.98.140402} {\bibfield  {journal} {\bibinfo  {journal}
  {Phys. Rev. Lett.}\ }\textbf {\bibinfo {volume} {98}},\ \bibinfo {pages}
  {140402} (\bibinfo {year} {2007})}\BibitemShut {NoStop}%
\bibitem [{\citenamefont {Jones}\ \emph {et~al.}(2007)\citenamefont {Jones},
  \citenamefont {Wiseman},\ and\ \citenamefont {Doherty}}]{WisemanPRA}%
  \BibitemOpen
  \bibfield  {author} {\bibinfo {author} {\bibfnamefont {S.~J.}\ \bibnamefont
  {Jones}}, \bibinfo {author} {\bibfnamefont {H.~M.}\ \bibnamefont {Wiseman}},
  \ and\ \bibinfo {author} {\bibfnamefont {A.~C.}\ \bibnamefont {Doherty}},\
  }\bibfield  {title} {\enquote {\bibinfo {title} {{Entanglement,
  Einstein-Podolsky-Rosen correlations, Bell nonlocality, and steering}},}\
  }\href {\doibase 10.1103/PhysRevA.76.052116} {\bibfield  {journal} {\bibinfo
  {journal} {Phys. Rev. A}\ }\textbf {\bibinfo {volume} {76}},\ \bibinfo
  {pages} {052116} (\bibinfo {year} {2007})}\BibitemShut {NoStop}%
\bibitem [{\citenamefont {Reid}(1989)}]{Reid}%
  \BibitemOpen
  \bibfield  {author} {\bibinfo {author} {\bibfnamefont {M.~D.}\ \bibnamefont
  {Reid}},\ }\bibfield  {title} {\enquote {\bibinfo {title} {{Demonstration of
  the Einstein-Podolsky-Rosen paradox using nondegenerate parametric
  amplification}},}\ }\href {\doibase 10.1103/PhysRevA.40.913} {\bibfield
  {journal} {\bibinfo  {journal} {Phys. Rev. A}\ }\textbf {\bibinfo {volume}
  {40}},\ \bibinfo {pages} {913--923} (\bibinfo {year} {1989})}\BibitemShut
  {NoStop}%
\end{thebibliography}%

\end{document}